\documentclass[manuscript]{emulateapj}
\usepackage{natbib}
\usepackage{times}
\usepackage{graphicx}
\usepackage{amsfonts,amsmath,amssymb}
\usepackage{latexsym}
\usepackage{mathrsfs}
\usepackage[usenames,dvipsnames]{color}
\usepackage{comment}

\newcommand{\cage}{$\langle$age$\rangle$} 
\newcommand{\cfeh}{$\langle$[Fe/H]$\rangle$}

\slugcomment{To appear in The Astrophysical Journal, Draft Version \today}
\shorttitle{Photometric ages and metallicities of GCs in the Virgo galaxy cluster core}
\shortauthors{Powalka et al.}

\begin{document}

\title{The Next Generation Virgo cluster Survey (NGVS). XXVI. The issues of photometric age and metallicity estimates for globular clusters}

\author{Mathieu Powalka\altaffilmark{1}, Ariane Lan\c{c}on\altaffilmark{1}, Thomas H. Puzia\altaffilmark{2}, Eric W. Peng\altaffilmark{3,4}, Chengze Liu\altaffilmark{5,13}, Roberto P. Mu\~noz\altaffilmark{2}, John P. Blakeslee\altaffilmark{6}, Patrick C\^ot\'e\altaffilmark{6}, Laura Ferrarese\altaffilmark{6}, Joel Roediger\altaffilmark{6}, R\'uben S\'anchez-Janssen\altaffilmark{7}, Hongxin Zhang\altaffilmark{2,3}, Patrick R. Durrell\altaffilmark{9}, Jean-Charles Cuillandre\altaffilmark{8}, Pierre-Alain Duc\altaffilmark{1,8}, Puragra Guhathakurta\altaffilmark{10}, S. D. J. Gwyn\altaffilmark{6}, Patrick Hudelot\altaffilmark{13}, Simona Mei\altaffilmark{11,12,16}, Elisa Toloba\altaffilmark{15}}

\affil{$^1$ Observatoire Astronomique de Strasbourg, Universit\'e de Strasbourg, CNRS, UMR 7550, 11 rue de l'Universit\'e, F-67000 Strasbourg, France}
\affil{$^2$ Institute of Astrophysics, Pontificia Universidad Cat\'olica de Chile, 7820436 Macul, Santiago, Chile}
\affil{$^3$ Department of Astronomy, Peking University, Beijing 100871, China}
\affil{$^4$ Kavli Institute for Astronomy and Astrophysics, Peking University, Beijing 100871, China}
\affil{$^5$ Center for Astronomy and Astrophysics, Department of Physics and Astronomy, Shanghai Jiao Tong University, Shanghai 200240, China}
\affil{$^6$ National Research Council of Canada, Herzberg Astronomy and Astrophysics Program, Victoria, BC, V9E 2E7, Canada}
\affil{$^{7}$UK Astronomy Technology Centre, Royal Observatory Edinburgh, Blackford Hill, Edinburgh, EH9 3HJ, UK}
\affil{$^8$ AIM Paris Saclay, CNRS/INSU, CEA/Irfu, Universit\'e Paris Diderot, Orme des Merisiers, 91191 Gif sur Yvette cedex, France}
\affil{$^9$ Department of Physics and Astronomy, Youngstown State University, Youngstown, OH, USA}
\affil{$^{10}$ UCO/Lick Observatory, University of California, Santa Cruz, 1156 High Street, Santa Cruz, CA 95064, USA}
\affil{$^{11}$ LERMA, Observatoire de Paris, PSL Research University, CNRS, Sorbonne Universit\'es, UPMC Univ. Paris 06, F-75014 Paris, France}
\affil{$^{12}$ Universit\'{e} Paris Denis Diderot, Universit\'e Paris Sorbonne Cit\'e, 75205 Paris Cedex13, France}
\affil{$^{13}$ Institut d'Astrophysique de Paris, UMR 7095 CNRS \& UPMC, 98bis Bd Arago, F-75014 Paris, France}
\affil{$^{14}$ Shanghai Key Lab for Particle Physics and Cosmology, Shanghai Jiao Tong University, Shanghai 200240, China}
\affil{$^{15}$ University of the Pacific, Department of Physics, 3601 Pacific Avenue, Stockton, CA 95211, USA}
\affil{$^{16}$ Jet Propulsion Laboratory, Cahill Center for Astronomy \& Astrophysics, California Institute of Technology, 4800 Oak Grove Drive, Pasadena, California, USA}

\email{Email: mathieu.powalka@astro.unistra.fr}

\begin{abstract}
Large samples of globular clusters (GC) with precise multi-wavelength photometry are becoming increasingly available and can be used to constrain the formation history of galaxies. We present the results of an analysis of Milky Way (MW) and Virgo core GCs based on five optical-near-infrared colors and ten synthetic stellar population models. For the MW GCs, the models tend to agree on photometric ages and metallicities, with values similar to those obtained with previous studies. When used with Virgo core GCs, for which photometry is provided by the Next Generation Virgo cluster Survey (NGVS), the same models generically return younger ages. This is a consequence of the systematic differences observed between the locus occupied by Virgo core GCs and models in panchromatic color space. Only extreme fine-tuning of the adjustable parameters available to us can make the majority of the best-fit ages old. Although we cannot exclude that the formation history of the Virgo core may lead to more conspicuous populations of relatively young GCs than in other environments, we emphasize that the intrinsic properties of the Virgo GCs are likely to differ systematically from those assumed in the models. Thus, the large wavelength coverage and photometric quality of modern GC samples, such as used here, is not by itself sufficient to better constrain the GC formation histories. Models matching the environment-dependent characteristics of GCs in multi-dimensional color space are needed to improve the situation.
\end{abstract}

\maketitle

\section{Introduction}
To study the star formation history of the universe, we may either measure star forming activity at all redshifts, or examine the present-day relics of past star formation activity and put them in temporal order. Numerous such relics come in the shape of globular clusters (GCs), whose relatively homogeneous stellar populations have, to first order, evolved passively since the time at which they formed. 

In the now dominant picture, metal-poor GCs formed mostly at high redshift in clumpy precursors of galaxies or in dwarf galaxies, possibly as early as the epoch of reionization \citep{cen2001,spitler2012}; while GCs with a variety of metallicities formed over a broader range of ages, from pre-enriched material in galaxies of various masses or during gas-rich galaxy merger events \citep{ashman1992, whitmore2000, li2014,renaud2017}. In such a scenario, the GC population of a large galaxy in the local universe results from a combination of early GC formation, the accretion of GCs from in-falling galaxies, and in-situ formation of GCs during gas-rich mergers \citep{cote1998, kisslerpatig1998, ferrarese2016}. And the most massive GCs are tentatively associated with star formation episodes in major mergers \citep{renaud2015} or alternatively with stripped compact cores of in-falling galaxies 
\citep[][and references therein]{liu2015}. If measured accurately enough, GC ages and chemical compositions can be used to test this picture, and thus provide fundamental information about the hierarchical assembly of galaxies. 

The most robust ages and abundances come from studies of individual stars or of resolved, decontaminated color-magnitude diagrams of star clusters, which are available only for clusters in the Milky Way (MW) and in the Local Group \citep{perina2009,dotter2011,beerman2012,leaman2013,vandenberg2013}. Studies of GCs in such local environments have found metal-poor objects to be old, and established that more metal-rich ones may have old or intermediate ages, with examples as young as $\sim$6\,Gyr \citep{marinfranch2009, forbes2010, dotter2011}. These findings have set the foundation of the theoretical picture sketched above. But it remains unclear to what extent such results are representative of the universe as a whole.

To reach further out, we must rely on the integrated cluster light. Unfortunately, spectra with  signal-to-noise ratios sufficient for high quality spectroscopic index measurements or for full-spectrum fitting techniques, remain difficult to obtain in large numbers beyond $\sim$10\,Mpc even with 10m class telescopes \citep[see][for examples]{puzia2005b, usher2015}. But they hold the promise of great diagnostic power at low and intermediate spectral resolutions \citep[e.g.][]{puzia2006, colucci2009, colucci2012, colucci2017, schiavon2013}. On the other hand, photometric surveys are producing rapidly growing catalogs of GC measurements in a broad variety of environments \citep[e.g ACSVCS, ACSFCS, NGVS, or SLUGGS respectively][]{cote2004,jordan2007,ferrarese2012,brodie2014}. Not only are the numbers of candidate GCs rising; the precision of the measurements has also improved to typically a few percent, and the accuracy to better than 5 percent \citep{powalka2016}. In this context, it is useful to provide an update on the analysis of GC colors with Evolutionary Population Synthesis (EPS) models. The issues we highlight in this update are relevant also when using colors to estimate luminosity-weighted ages or metallicities of whole galaxies, or when determining photometric redshifts.

GC color distributions are the basis of a vast body of literature that established strong relations between the empirical properties of GCs and their environment \citep[e.g][]{peng2006,powalka2016L}. A major challenge of these studies is the conversion of color distributions into metallicity distributions. Two main hurdles in this exercise are the age-metallicity degeneracy of broad-band colors and the model-dependence of their interpretation. The precise conversion of colors into metallicity requires prior ideas about age \citep{geisler1996}. Most studies discuss this problem but then adopt one or a few color-metallicity relations based on EPS model predictions at old ages, or based on old stellar populations in the Local Group. This approach may be unable to capture the real complexity of remote GC populations. A clear manifestation of model-dependence issues is found in the debate on bimodality: while GC color distributions tend to be bimodal at least around massive galaxies, it remains unsettled whether or not this implies a true bimodality of the metallicity distribution \citep[][]{yoon2011,blakeslee2012,chiessantos2012}.

Estimating GC ages from colors is even more difficult than evaluating metallicities, and as a result there are only few attempts in the recent literature. In principle, the combined usage of near-UV, optical and near-IR fluxes should provide a better handle on age than optical colors alone \citep{puzia2002, hempel2004,anders2004,kotulla2009}: the optical-infrared colors are sensitive to metallicity through the temperature of the red giant branch, and near-UV to optical colors then measure the properties of the turn-off, hence age. A first difficulty arises from the possible existence of blue stragglers, hot sub-giants or extended horizontal-branch stars in non-standard proportions. An excess of any of these types of stars can make the short-wavelength colors of stellar populations bluer, leading to artificially young age estimates \citep[][]{cenarro2008,koleva2008,xin2011,chiessantos2011}. A second, and equally frustrating difficulty comes again from the model-dependence of theoretical color predictions. In color-color diagrams, the loci predicted by any given population synthesis code for GCs of various old ages and metallicities are generally narrow. When age and metallicity are not fully degenerate, age tends to be related to the position across this narrow locus. Small systematic errors in either the data or the models, if they are oriented across to the main locus, strongly affect age estimates. When different models predict mutually exclusive loci in parts of color-color diagrams, the situation becomes rather confused.

In this article, we re-assess GC age and metallicity estimates
in the context of recent high-quality multi-band photometric measurements. Our main working sample consists of $\sim 2 \times 10^3$ luminous GCs located in the central four square degrees of the Virgo galaxy cluster (i.e. within $\sim 300$\,kpc projected distance from M87), that were identified as part of the Next Generation Virgo cluster Survey (NGVS; \citealt{ferrarese2012}) and its near-infrared counterpart NGVS-IR \citep{munoz2014}. In the first paper of this series \citep[][hereafter Paper I]{powalka2016}, we provided accurately calibrated $u^*grizK_s$ magnitudes for these objects. The combination of near-UV, optical and near-IR data, together with strict limits on photometric errors and spatial extent, ensure low contamination by stars or by background galaxies. 
We showed that the GCs in the Virgo core define a very narrow locus in
the ($u-i$)\,vs.\,($i-K_s$) color-color diagram, but reveal sequences with more substructure in some other color-color combinations. In a subset of diagrams, such as ($g-r$) vs. ($i-z$), the GCs most centrally located (i.e. in M87) separate from many of those located further out \citep{powalka2016L}. On average, those located centrally have redder ($i-z$) colors at a given ($g-r$)\footnote{Alternatively, these GCs show bluer ($g\!-\!r$) colors at a given ($i\!-\!z$) color.} than those located further out. In these diagrams, we also showed that the colors of MW globular clusters resemble those of some of the more external Virgo GCs, rather than those of the clusters inside M87. These findings illustrate that the environmental dependance of GC colors cannot be summarized with just one parameter, because all colors should then vary in unison. 

Our aim is to examine what generic trends in terms of age and metallicity result from the comparison of the Virgo core and Milky Way GC colors with standard EPS models. The formal age and metallicity estimates in this article are the typical values one may obtain using a combination of four/five colors from the near-UV to the near-IR, and ten synthetic single stellar population (SSP) models. As we will show, generic trends emerge despite the variance between individual models. We then confront these generic results with the current picture of GC formation and discuss the consequences, keeping in mind that not all the best fits are actually good fits of the GC spectral energy distributions, and that standard EPS models make a number of simplifying assumptions.

The article is organized as follows. In Sections~\ref{sec_data} and \ref{sec_models}, we respectively present the GC samples and the SSP models that we are using in this work.
In Section~\ref{defin}, we describe the methods implemented to estimate the ages and metallicities of the GCs.
~It leads to Section~\ref{sec:results} where we apply our method first to Milky Way objects, then to the Virgo core GC sample, with an analysis of the results. Section~\ref{sec:discussion} is devoted to a discussion of the caveats and of the consequences of these results. In Section~\ref{conclusion} we conclude with an overview of our findings.

\section[]{The data}
\label{sec_data}
The first set of data used in this paper is taken from the NGVS pilot field sample of Paper~I\footnote{Data available at: http://cdsarc.u-strasbg.fr/viz-bin/Cat?J/ApJS/227/12} .
It consists of 1846 bright GCs in the central 3.62~$\rm{deg}^2$ of the Virgo galaxy cluster, 
observed in the $u^*$, $g$, $r$, $i$, $z$ and $K_s$ filters (MegaCam/WIRCam filters
of the 4-meter Canada France Hawaii Telescope).
These GCs are selected using aperture corrected magnitudes in the $uiK$ diagram ($u-i$ versus $i-K_s$) which 
significantly reduces the possible contamination by foreground stars or background galaxies \citep{munoz2014}.
In addition, the sample is restricted to objects with {\sc SExtractor}\footnote{\citet{bertin1996}} 
magnitude errors smaller than 0.06 mag in each band.
 
Paper~I lists aperture corrected magnitudes for two alternative flux calibrations. 
Following the recommendations provided there, 
we adopt the empirical calibration against SDSS \citep[Sloan Digital Sky Survey DR10, see][]{ahn2014} 
and UKIDSS stars \citep[UKIRT Infrared Deep Sky Survey, see][]{lawrence2007, casali2007}, rather than the 
calibration against synthetic stellar photometry. We also apply the systematic corrections to $u^*$ and $z$ 
recommended by the SDSS DR10 flux calibration pages\footnote{https://www.sdss3.org/dr10/algorithms/fluxcal.php},
as suggested in the discussion of systematic errors in Paper~I (i.e.~we apply a shift of $-0.04$\,mag 
to $u^*$ magnitudes and a shift of $0.02$\,mag to $z$ magnitudes). All colors and fluxes used
in the following  are dereddened.  
For more details on the data selection, the individual photometric errors, the extinction corrections, 
a budget of systematic errors, and an overall description of this sample we refer the reader to Paper~I.
All in all, it produces a large and well-understood GC photometric sample.
 
The strict selection criteria of the sample remove low-mass GCs ($\la\! 10^{5.5}\,M_{\odot}$),
with a mass cut-off that depends on the GC metallicity and age. 
In particular, the limits on $u^*$ errors restrict the mass range at the red end of the GC color distribution, compared to the blue end.
~The GC selection function is complex, and we postpone corrections for completeness to future studies of a larger area in Virgo.
~Therefore, the catalog only contains bright/massive Virgo GCs, and we defer the detailed studies of inferred distributions, such as discussions of the metallicity bimodality, to subsequent papers.
\smallskip
 
The Milky Way sample is selected from the target sample of the Panchromatic 
High-Resolution Spectroscopic Survey of Local Group Star Clusters (NGC\,104, NGC\,288, NGC\,362, NGC\,1851, NGC\,1904,
NGC\,2298, NGC\,2808, NGC\,6656, NGC\,7078, NGC\,7089,
and NGC\,7099; \citealt{schonebeck2014}). It consists of VLT/X-
shooter spectra of MW GCs, accurately calibrated in absolute flux ($\pm 2.5 \%$) and covering wavelengths from the near-UV to near-IR.
As detailed in \citet{powalka2016L}, we derive the MegaCam $u^*griz$ magnitudes using the transmission curves of \citet{betoule2013}.
The dereddening is done using the extinction values of the McMaster catalog (\citealt[][]{harris2010}; based on \citealt{webbink1985,zinn1985,reed1988}).
For any additional information, we refer to \citet{schonebeck2014} and future papers.

\section[]{The models}
\label{sec_models}
\begin{deluxetable}{lcr}
\tablecaption{\label{tab:ssp} Stellar libraries and isochrone references for the different SSP models used in this paper. }
\tablehead{
\colhead{Model} 	& \colhead{Stellar library**} 	& \colhead{Isochrones} 	
}
\startdata
BC03 & STELIB & Padova 1994 \\
BC03B & BaSeL 3.1 & Padova 1994   \\
C09BB & BaSeL 3.1 & BaSTI \\
C09PB & BaSeL 3.1 & Padova 2007 \\
C09PM & MILES & Padova 2007 \\
M05 & BaSeL 3.1 & Cassisi \\
MS11 & MILES & Cassisi \\
PEG & BaSeL 2.2 & Padova 1994 \\
PAD & ATLAS ODFNEW / PHOENIX BT-Settl & PARSEC 1.2S  \\
VM12 & MILES* & Padova 2000 \\ 
\enddata
\tablecomments{ATLAS ODFNEW refers to \citet{castelli2004}. BaSeL: \citet{lejeune1997, lejeune1998, westera2002}. BaSTI: \citet{pietrinferni2004,cordier2007}. Cassisi: \citet{cassisi1997a, cassisi1997b, cassisi2000}. MILES: \citet{sanchezblazquez2006}. Padova 1994: \citet{alongi1993, bressan1993, fagotto1994a, fagotto1994b, girardi1996}. Padova 2007: \citet{girardi2000, marigo2007, marigo2008}. PARSEC 1.2S: \citet{bressan2012, tang2014, chen2014, chen2015}. PHOENIX BT-Settl: \citet{allard2003}. STELIB: \citet{leborgne2003}.
(*) The MILES library extends from 3464 to 7500\,\AA. In order to reach $u^*$, $i$ and $z$ magnitudes, we used the combinaison of NGSL and MIUSCAT from \citet[][ they provide wavelengths from $\sim$1700\,\AA\ to $\sim$9500\,\AA]{koleva2012} and we extrapolated the NGSL+MILES+MIUSCAT spectra from 9500\,\AA\ to 10000\,\AA\ using Pegase GC spectra.
(**) Optical-only libraries are generally combined with BaSeL at UV and IR wavelengths in the original codes.}
\end{deluxetable}

In Paper~I, we compared the locus of the GC sample in various two-color planes with the predictions from seven commonly used EPS codes for simple stellar populations (SSP) at old ages. Only EPS codes with broad age-metallicity coverage and with a range of wavelengths adequate for synthetic $u^*griz(K_s)$ photometry, were included.
~Based on these codes, we produce ten grids of synthetic SSP colors for the present study, with a more extended age range than in Paper~I.

\begin{figure*}[!t]
\centering
\includegraphics[width=1.0\linewidth]{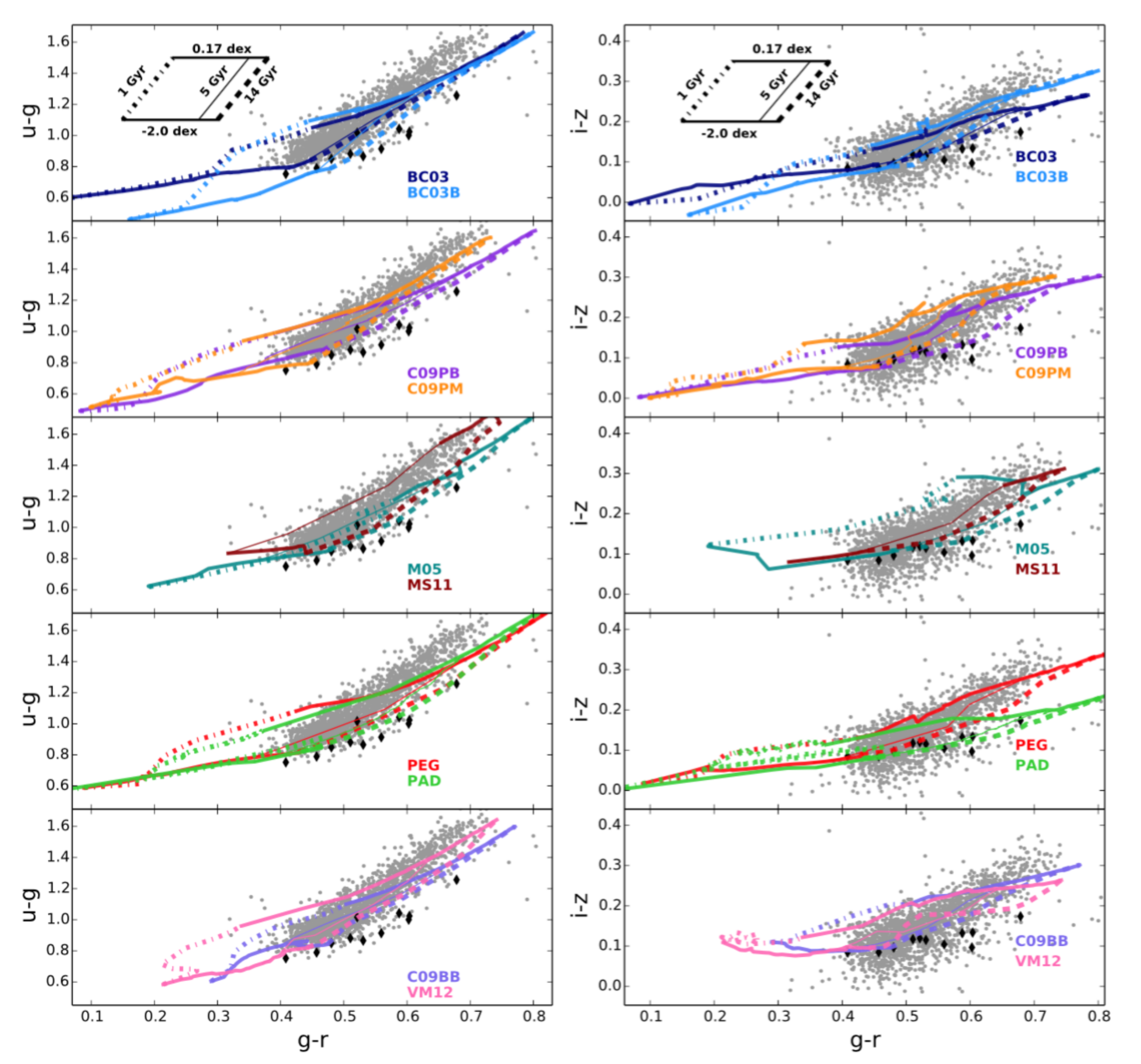}
\caption{\label{ccd_comparison} Color-color comparison of the Virgo core GCs of Paper~I with the 10 SSP model grids used in this article. The grids extend from Age\,=\,1\,Gyr to 14\,Gyr (instead of 6\,Gyr to 13\,Gyr for Paper~I) and from [Fe/H]\,=\,-2.0\,dex to 0.17\,dex. The only exception is the MS11 grid which starts at 5\,Gyr. The NGVS GCs are shown in gray whereas the black diamonds refer to the MW GCs used in this paper}.
\end{figure*}

The first two are obtained with the EPS code of \cite{bruzual2003} and are named {\bf BC03} and {\bf BC03B}.
The next three, {\bf C09PB}, {\bf C09BB} and {\bf C09PM} are computed with the Flexible Stellar Population Synthesis (FSPS v2.6) of \citet{conroy2009}.
We also use a model from \citet{maraston2005} EPS code, labelled {\bf M05} and one from \citet{maraston2011}, named {\bf MS11}.
We embed one model from the CMD 2.8\footnote{\tt http://stev.oapd.inaf.it/cgi-bin/cmd} EPS code \citep{bressan2012} named {\bf PAD},
and one based on the PEGASE model \citep{fioc1997}\footnote{We use the code made available as {\sc Pegase-HR} by \citet{leborgne2004}.}, labelled {\bf PEG}.
Finally, we use the model detailed in \citet{vazdekis2012, ricciardelli2012}, labelled {\bf VM12}.
For convenience, we recall some of the SSP model properties and ingredients in Table\,\ref{tab:ssp}, and illustrate some of their predicted colors in Figure\,\ref{ccd_comparison} (see Paper~I for more diagrams).\\

Our default grid covers ages between 1 and 14\,Gyr with uniform linear sampling in time, and metallicities between 0.0002 and 0.03 
with uniform linear sampling in Z. However, two models are unable to cover this grid, namely C09BB and MS11. C09BB does not reach metallicities lower than 0.0003, whereas MS11 does not cover ages lower than 5\,Gyr.
We include these models in our computation but we verified that these different limits are not biasing our age-metallicity estimates (inferred ages do not pile up at the edges of these grids).
All model colors are computed with a \citet{kroupa1998} or \citet{kroupa2001} initial mass function (IMF) and calculated at the typical redshift of the Virgo cluster core ($z\,\simeq\,0.004$).
This is done, whenever possible, by measuring synthetic colors on redshifted spectra of SSP models.
For SSP models that provide only colors, we apply the $k$-corrections determined from the SSP model spectra of M05 and PEG.

In this paper, we convert between [Fe/H] and the metallicity Z with the approximate formula [Fe/H]~$=\!\log($Z$/0.02)$.

\section{The method}
\label{defin}

\subsection{Analysis with one Model Grid}

The method used in this paper is based on the comparison of observed GC colors and theoretical color predictions with the likelihood function as reference.

In practice, each grid of synthetic SSP colors is sampled at 100 age values between 1 and 14\,Gyr, and 550 [Fe/H] values between --2.0 and 0.17. These steps are chosen so that the mean photometric errors of the observed GCs are larger than the spacing between two neighbouring grid points.
We assume that the errors on magnitudes are approximately gaussian, and account for correlations between two colors that have a photometric passband in common. Hence the likelihood of any age and metallicity ($Z$) for a 
given cluster is written: 
$$
L ({\rm age}, Z)=\frac{1}{\sqrt{(2\,\pi)^n\,|\Sigma|}}\,e^{-\frac{1}{2}\,(x - \mu)^T\,\Sigma^{-1}\,(x - \mu)}
$$
where $x$ contains the colors of the cluster, $\mu\,=\,\mu ($age,\,Z$)$ contains the model colors for a given age and metallicity, the superscript $T$ denotes the transpose,  
$\Sigma$ is the covariance matrix of the random errors on the observed colors, and $n$ the number of colors used.
The maximum likelihood provides the adopted estimate of age and metallicity for the cluster.

We have considered Bayesian estimates of age and metallicity as an alternative to maximum-likelihood estimates.
In the Bayesian approach, the posterior probability of an age-metallicity pair is proportional to the product of $L($age,\,Z$)$ and the adopted prior probability $p_0$(age,\,Z).
The Bayesian estimates of age and metallicity are then the weighted averages of age and metallicity,
in which the weights are given by their posterior probabilities.
However, in the case of the GCs, the age/metallicity degeneracy frequently results in likelihoods with remote secondary maxima, which leads to meaningless averages, unless one provides strong priors.

We wish to avoid strong priors, because they would mask interesting trends in the 
model-data comparison that go beyond the well-known age-metallicity degeneracy.
Figure 1 shows that individual model grids are offset from the region occupied by Virgo core GCs in (model-dependent) parts of color-color space. For GCs located outside the area occupied by a model grid, selecting even the closest model in color-space via maximum-likelihood will not always provide a good fit. But it will allow us to expose the global direction of the offset between the data and the model grid, and the direction of the corresponding bias in age and metallicity. By setting a prior (for instance by choosing old ages, as was done in numerous GC studies in the literature) we would lose this information.

\subsection{Analysis with the Combined Model Grid Set}
\label{sec:ce}
Using ten SSP model grids, we obtain ten different age and metallicity estimates.
As we will see in Section\,\ref{sec:results}, the full width of the distributions of these ten estimates can be very large.
But closer examination shows that the full width of the broadest distributions is usually dominated by few outliers, the other estimates being significantly more internally consistent.
Which models produce outliers depends on each particular GC color in a complex way.

Based on this result, we have designed a ``concordance estimate" (CE) of the age and metallicity of any given GC.
This method averages the ten individual estimates using weights that effectively reject outliers.
Let $\langle\Delta {\rm Age}\rangle_i$ be the average of the differences between the age obtained with the {\it i}-th model and the age obtained with the other models.
$\langle\Delta {\rm [Fe/H]}\rangle_i$ is the analog for the estimated metallicities. For the GC under study, we write the weight of the {\it i}-th model:
\begin{equation*}
 w_i = (w_{i,{\rm Age}} \times w_{i,{\rm [Fe/H]}})\,\times\,\frac{L^{max}_{i}}{\sum_{j=0}^{N_{model}}\,L^{max}_{j}}
\end{equation*}
In this expression, $w_{i,{\rm Age}}$ takes a value of 1 if $\langle\Delta {\rm Age}\rangle_i$ is smaller than 3\,Gyr and 0 if $\langle\Delta {\rm Age}\rangle_i$ is larger than 7\,Gyr,
and drops linearly from 1 to 0 between these thresholds. The shape of $w_{i,{\rm [Fe/H]}}$ is similar, with thresholds at $\langle\Delta {\rm [Fe/H]}\rangle_i$ of 0.4 and 1 dex. $L^{max}_{i}$ is the maximum likelihood of the {\it i}-th model.
We note that using a simpler median, or a
more complex iterative outlier rejection, would lead to similar results, i.e. ages within about 2 Gyr of those we retained, and metallicities within 0.3 dex.

The CE of age and metallicity, which we write as \cage\ and \cfeh, represent the photometric estimates the ten models we have considered tend to agree upon.
They may still be severely biased, for instance if the model predictions are inappropriate for a common reason (e.g. inadequate chemical composition of the GC constituent stellar populations, unusual stellar mass occupation functions of the contributing stellar evolutionary phases, etc.).

It is worth mentioning that we do not use a weighting scheme based on a global ranking of the models.
~For instance, such weights could be based on Bayes evidence ratios, that measure the relative ability of any model to reproduce the set of observed GC colors as a whole.
~The Bayes ratios (which we did compute) are very far from one  and would in effect select a single model.
~But which model is preferred depends sensitively on $i)$ the particular age/metallicity grid size adopted (i.e.~whether or not young ages or extreme metallicities are included),
$ii)$ the GC subsample characteristics (i.e.~whether there are more red or more blue clusters),
and $iii)$ the set of photometric passbands used.
~As shown in Paper~I, none of the models available to us provides a statistically acceptable representation of the observed color locus as a whole, and deserves to be strongly predominant.
~As a consequence, we prefer a concordance estimate based on the internal consistency of the model set for the particular SED coverage of our data.

\section{Results}
\label{sec:results}
\subsection{Milky Way Globular Clusters}
\label{sec:MWresults}

\begin{figure*}[!t]
\centering
\includegraphics[width=0.8\linewidth]{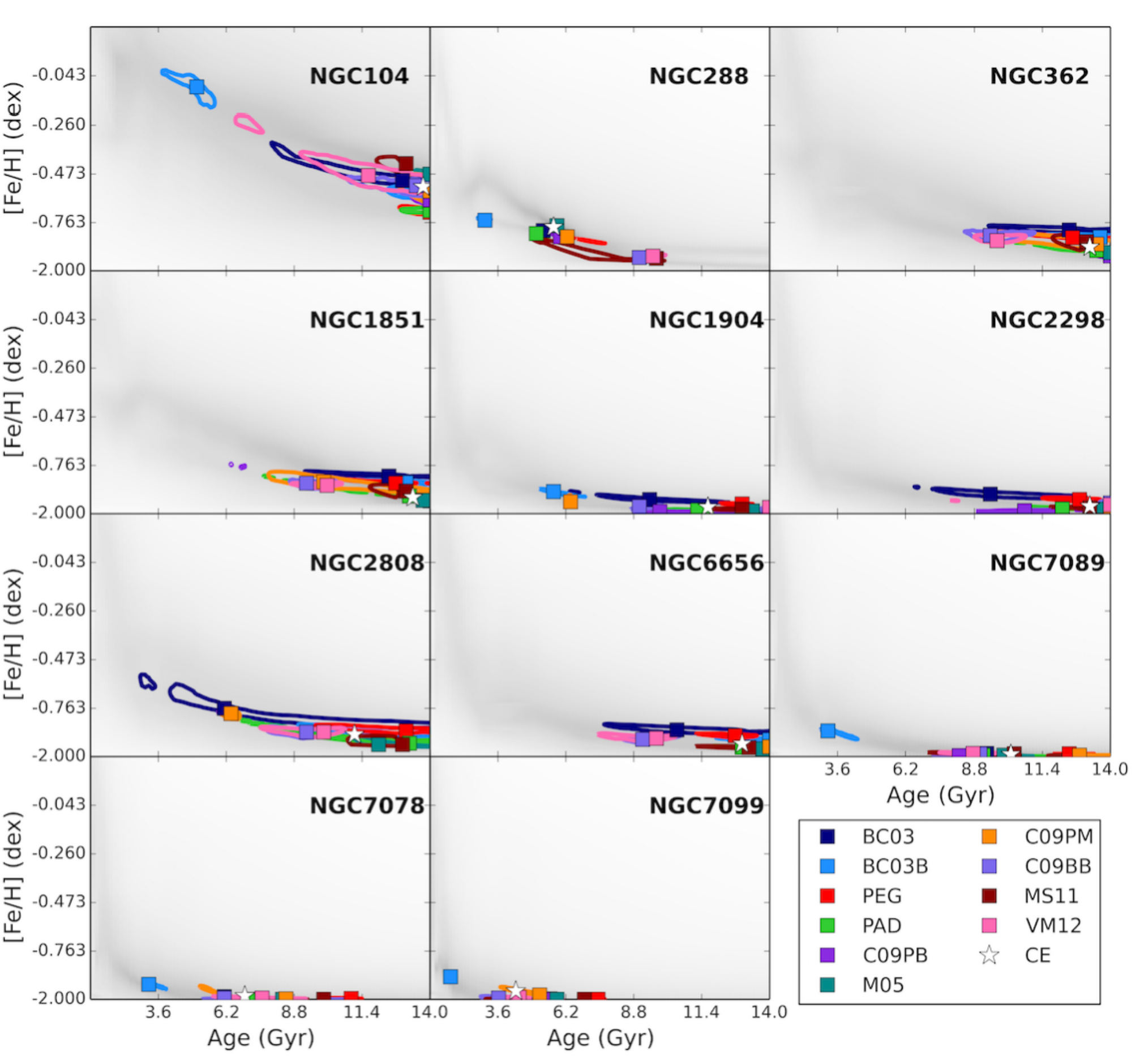}
\caption{\label{MWGCbanana} Likelihood distributions in the age-metallicity plane for each of the Milky Way GCs as obtained with the ten SSP models (see Tab.\,\ref{tab:ssp}), using $u^*griz$ photometry.
The contour levels are set to 90\,\% of the maximum likelihood. The CE is highlighted with a white star. The shaded background shows the sum of the likelihoods obtained with the ten SSP model grids.}
\end{figure*}
 
\begin{deluxetable*}{llccccl}[b]
\tablecaption{\label{tab:MWGCs} MW GC age and metallicity estimates. The errors on age and [Fe/H] only reflect the discrepancies between the models.}
\tablehead{
  \colhead{NGC} &  \colhead{Name} & \colhead{\cfeh} & \colhead{$<$Age$>$} & \colhead{Age$_{\rm lit.}$}  & \colhead{[Fe/H]$_{\rm lit.}$} & \colhead{Refs.}\\
                           &                             & \colhead{[dex]} & \colhead{[Gyr]}  & \colhead{[Gyr]}& \colhead{[dex]}   & 
}
\startdata
104   &  47 Tuc & $-0.53\pm 0.16$ & $13.64\pm 2.59$ & $12.75\pm0.50$ & $-0.76\pm0.02$ & (1),(2)\\
288   &              & $-0.80\pm 0.19$ & $  5.77\pm 2.01$ & $12.50\pm0.50$ & $-1.32\pm0.02$ & (1),(2)\\
362   &              & $-1.04\pm 0.12$ & $13.12\pm 1.58$ & $11.50\pm0.50$ & $-1.30\pm0.04$ & (1),(2)\\
1851 &              & $-1.18\pm 0.14$ & $13.30\pm 1.62$ &   $9.20\pm1.10$ & $-1.18\pm0.08$ & (1),(3)\\
1904 &              & $-1.49\pm 0.21$ & $11.61\pm 2.75$ & $11.70\pm1.30$ & $-1.58\pm0.02$ & (1),(3)\\
2298 &              & $-1.43\pm 0.17$ & $13.14\pm 1.50$ & $13.00\pm1.00$ & $-1.96\pm0.04$ & (1),(2)\\
2808 &              & $-1.07\pm 0.20$ & $11.12\pm 2.87$ & $9.30\pm1.10$ & $-1.18\pm0.04$ & (1),(3)\\
6656 &      M22 & $-1.26\pm 0.16$ & $12.86\pm 1.78$ & $12.30\pm1.20$ & $-1.70\pm0.08$ & (1),(3)\\
7078 &      M15 & $\leq -2$         & \nodata                & $13.25\pm1.00$ & $-2.33\pm0.02$ & (1),(2)\\
7089 &        M2 & $-1.74\pm 0.24$ & $  10.17\pm 2.47$ & $12.50\pm0.75$ & $-1.66\pm0.07$ & (1),(2)\\
7099 &      M30 & $\leq -2$               & \nodata                & $13.25\pm1.00$ & $-2.33\pm0.02$ & (1),(2)

\enddata
\tablecomments{Metallicity: (1)~\citet{carretta2009}; Age: (2)~\cite{dotter2010}; (3)~\cite{salaris2002}.}
\end{deluxetable*}

To assess our stellar population characterization method in a well studied context, we begin with the analysis of the colors of Milky Way GCs,
for which the literature provides reference parameters,
and for which synthetic $u^*$, $g$, $r$, $i$, $z$ photometry in the passbands of the NGVS survey is available from flux-calibrated VLT/X-shooter spectra \citep[see Paper~I;][]{schonebeck2014, powalka2016L}.
~We obtain maximum-likelihood and concordance estimates using the colors ($u\!-\!g$), ($g\!-\!r$), ($r\!-\!i$), and ($i\!-\!z$).
We present in Figure~\ref{MWGCbanana} the age-metallicity likelihood analysis with each of the ten SSP models, for every Milky Way GC.
~Every SSP model grid gives somewhat different results along with the noticeable degeneracy between age and metallicity illustrated by the shape of the confidence contours.

We summarize the derived \cage\ and \cfeh\ values in Table\,\ref{tab:MWGCs} together with the literature values.
~The error bars on our estimates reflect only the discrepancies between the models in the framework of the CE values.
~These uncertainties are computed as the root mean square deviation of individual estimates from the CE.
~We find generally reasonable agreement between the results of our SSP model inference method and the literature results, which are based on various techniques such as optical and near-IR isochrone-fitting,
full-spectrum fitting, RR Lyrae star variability, etc.
~The GCs are found to be old, with a range of low metallicities. A notable exception is NGC\,288 for which the CE age is significantly younger than the literature value.
This GC is known to have a well populated blue horizontal branch (HB) \citep{lee1994} that could be 
responsible for this seemingly young age. However, NGC\,1904, NGC\,2298, and NGC\,7089 also share a similar HB morphology 
without this leading to a young inferred age.
~Two of the Milky Way GCs (NGC\,7078 and NGC\,7099) have literature metallicities lower than $-2.0$ dex, which are outside the formal limits of our SSP models.
We keep those GCs in the sample, but flag them as less reliable in their \cage\ and \cfeh\ estimates.

\subsection{Virgo Core Globular Clusters}
\label{sec:VirgoGCresults}

\begin{figure*}
\centering
\includegraphics[width=0.70\linewidth]{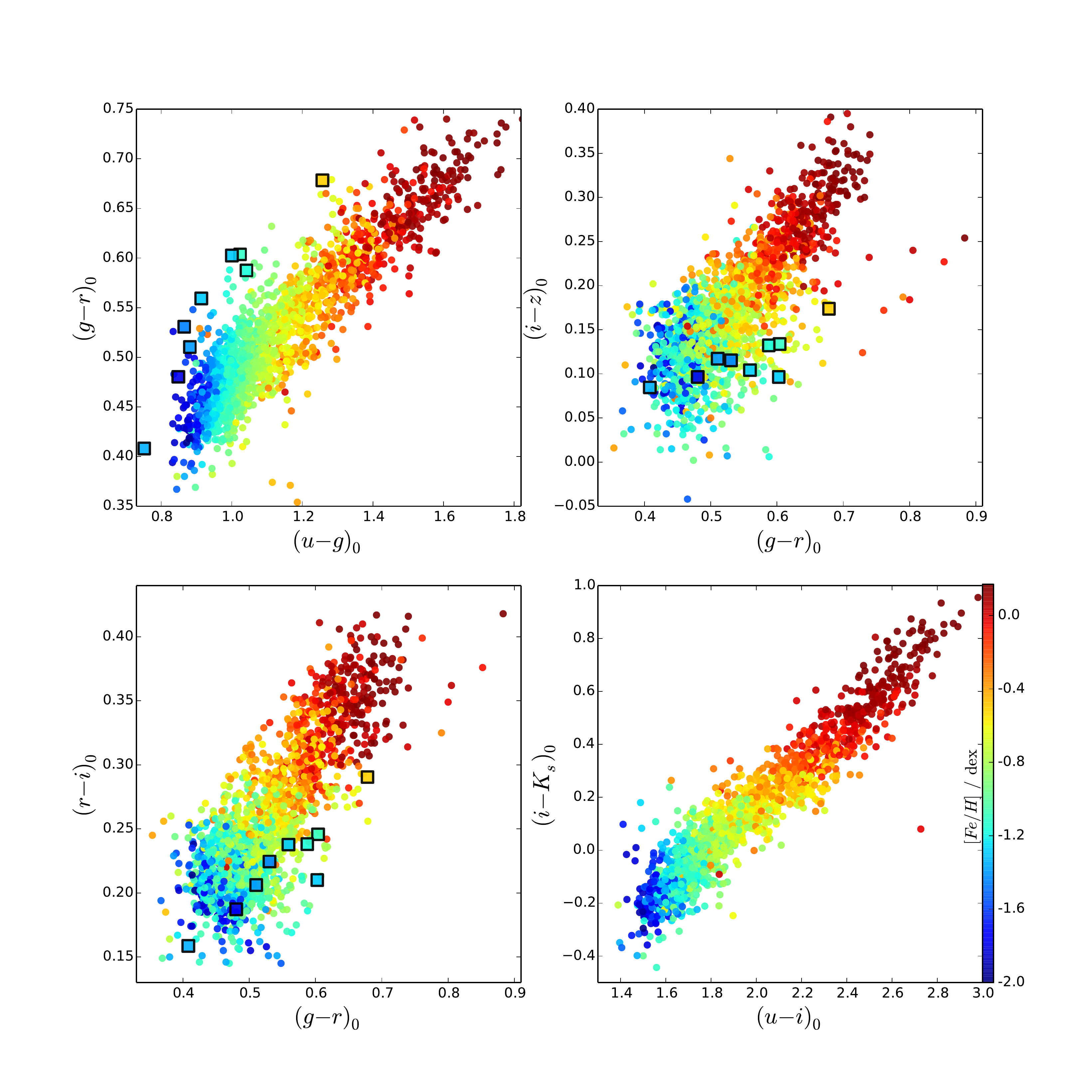}
\caption{\label{VirgoGCcolfeh} Color-color diagrams for Virgo (dots) and MW GCs (squares) with their concordance metallicity, \cfeh, encoded by their symbol colors.
As we do not have the $K_s$ band for the MW GCs, the $uiK$ diagram is showing solely the Virgo GCs.}
\end{figure*}

\begin{figure*}
\centering
\includegraphics[width=0.70\linewidth]{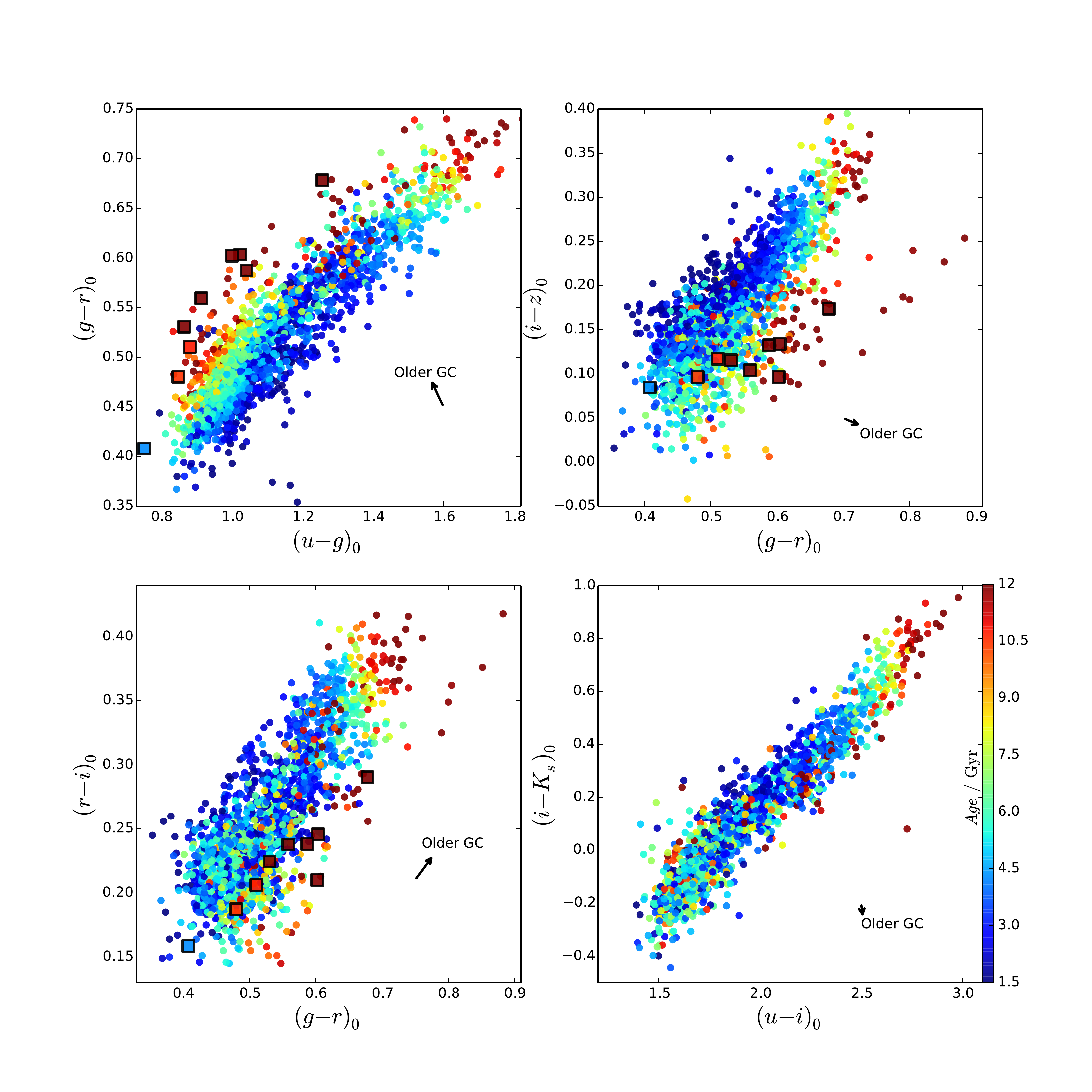}
\caption{\label{VirgoGCcolage} Color-color diagrams for Virgo (dots) and MW GCs (squares) with their concordance age, \cage, encoded by their symbol colors.
As we do not have the $K_s$ band for the MW GCs, the $uiK$ diagram is showing solely the Virgo GCs. 
The black arrow shows the direction in which a shift of the observed colors would maximize the number of old ages the concordance estimate
returns (the amplitude of the plotted vector is five times the 1-sigma
shift described in Section\,\ref{sec:ZPerrors}).}
\end{figure*}

Applying the same method to the $u^*griz$ colors of the Virgo core GCs,
we obtain their concordance age and metallicity estimates and illustrate their dynamic ranges in the color-color planes shown in Figures~\ref{VirgoGCcolfeh} and \ref{VirgoGCcolage}.
~We note that, in general, the diagrams exhibit a smooth scaling of GC colors with concordance age and metallicity, however, with a noticeable degree of degeneracy between these parameters.
The $uiK$ diagram shows a continuous increase of assigned metallicity with redder colors 
along the principal direction of the data distribution.
At the same time, the color dynamic range in this diagram is almost entirely degenerate in age, 
which leads to a relatively narrow sequence of data compared to the other diagrams.
This property is what makes the $uiK$ diagram such an ideal tool for separating foreground stars, background galaxies, and star clusters,
and selecting GCs at Virgo-like distances with accurate photometry alone \citep[see][]{munoz2014, powalka2016}.
In contrast, the $ugr$ diagram highlights a distinct gradient of both age and metallicity. Therefore, this diagram turns out to be useful when deriving the age and metallicity of a GC.

Although the M87 core GCs are believed to have spectroscopic old ages \citep[$\sim$\,13\,Gyr;][see Appendix\,\ref{setvscohen} for a comparison]{cohen1998},
our model inference returns numerous ages younger than 6\,Gyr, while only few of the inferred 
ages are old. We will discuss this result extensively in later sections.

As discussed in \cite{powalka2016L}, the ($g\!-\!r$) vs.~($i\!-\!z$) plane displays significantly different relations for Virgo core and Milky Way GCs.
~A similar deviation is visible in the ($u-g$) vs. ($g-r$) plane.
~As expected from Section\,\ref{sec:MWresults}, the Milky-Way GCs are located in the region of color-color space in which our analysis leads to old ages.
~A subset of Virgo core GCs also populates that region, but the majority of the Virgo GCs (in particular those located closer to M87), are offset.
~This offset drives differences in derived ages and metallicities when both sequences are analyzed with identical models and methods.
~We will return to this point after having examined the behaviour of the parameters derived from individual model sets.

\subsubsection{Testing Model-to-Model Consistency}
\begin{figure*}
\centering
\includegraphics[width=0.8\linewidth]{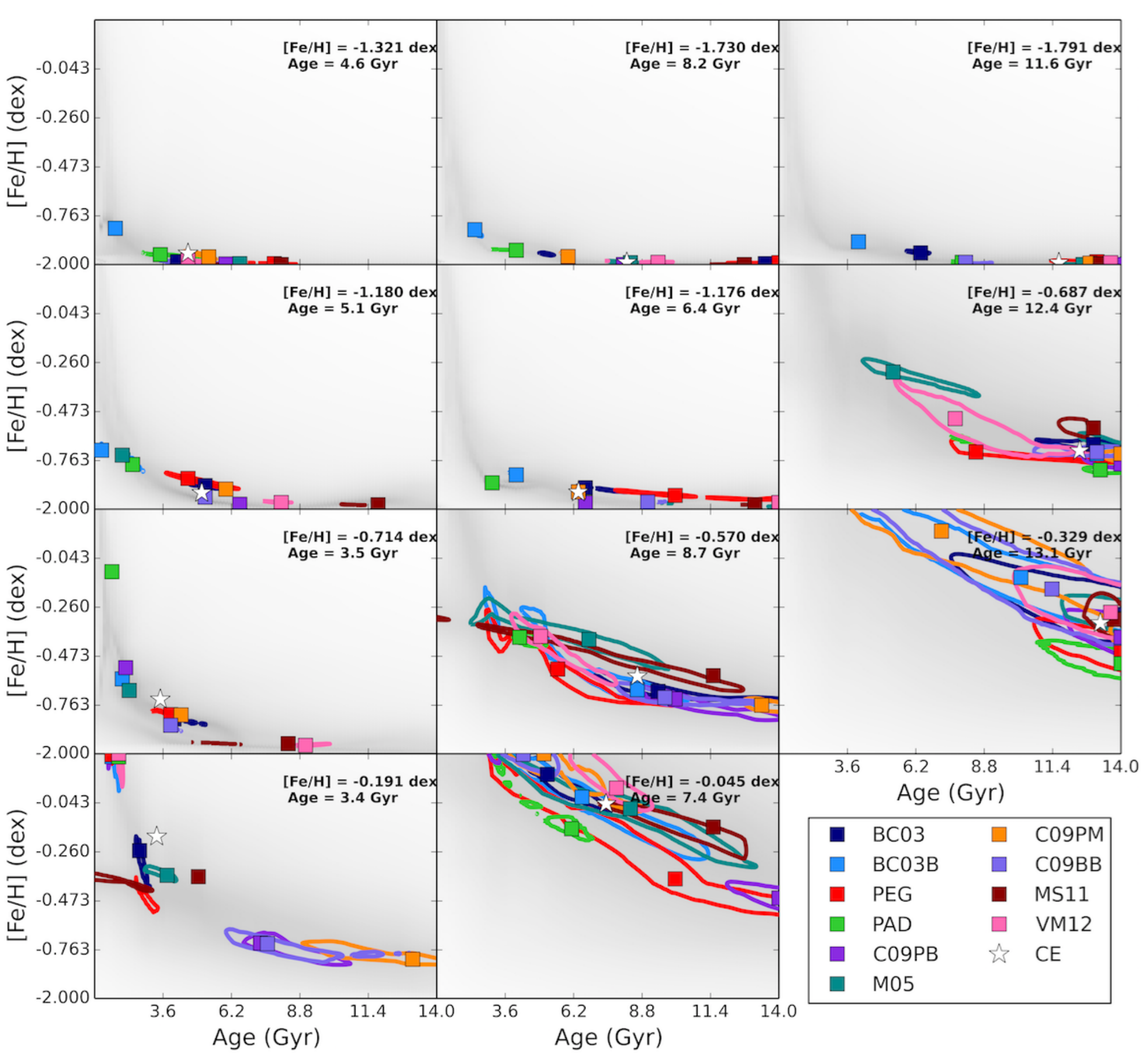}
\caption{\label{VirgoGCbanana} Likelihood distributions in the age-metallicity plane for eleven representative Virgo GCs obtained with the ten SSP models using $u^*griz$ photometry. The layout is as in Figure\,\ref{MWGCbanana}. }
\end{figure*}

\begin{figure*}[!t]
\begin{center}
\includegraphics[width=0.8\linewidth]{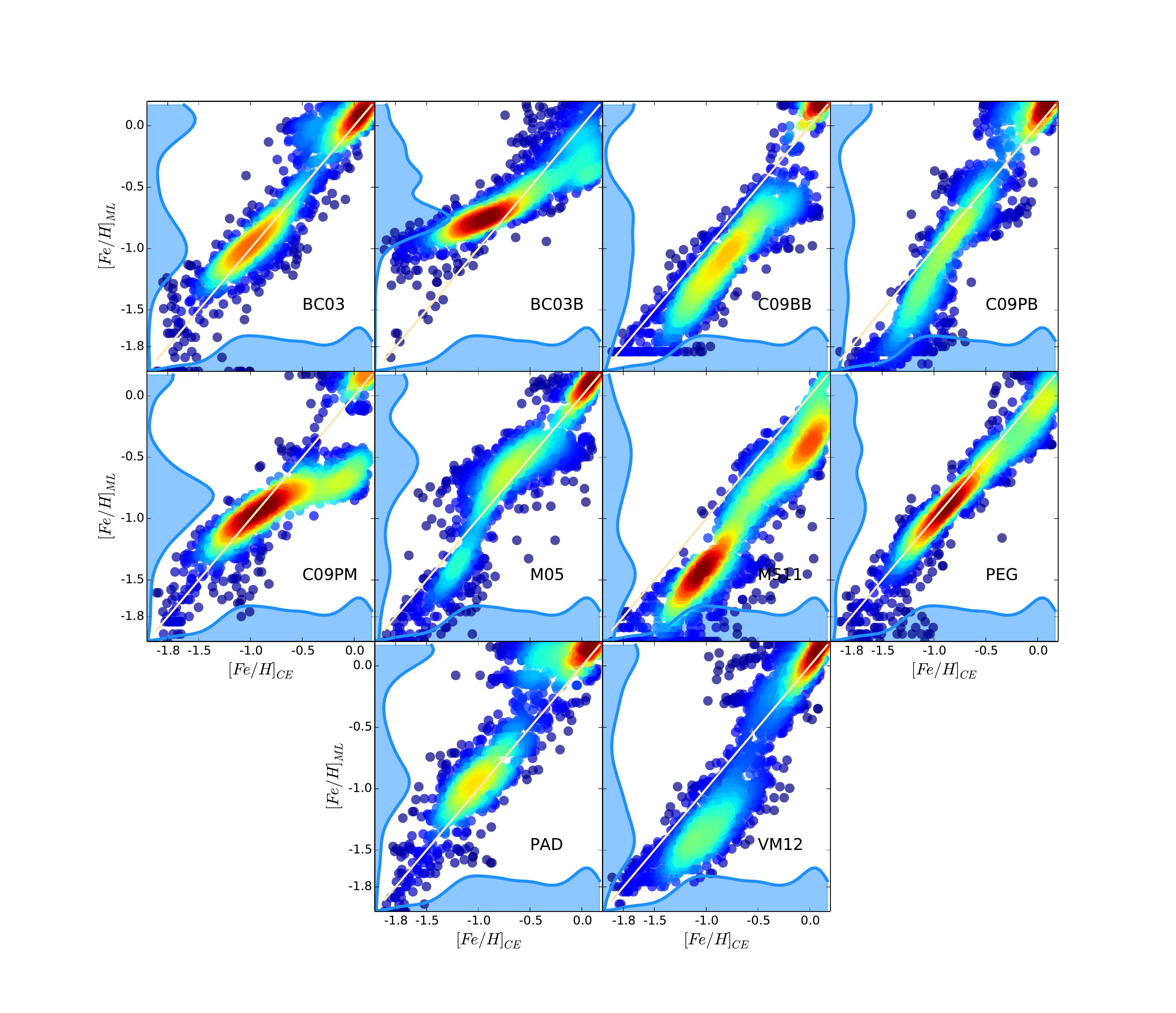}
 \caption{\label{fig:metmet} Metallicity distributions predicted by each SSP model (y-axis) compared with the CE (x-axis) for our entire Virgo core GC sample. The color shading of the symbols indicates the local data density, with dark red showing the highest concentrations.
 We show in beige the one-to-one line which highlights if a model is consistent with the CE i.e with a weighted mean of the other models. Each panel shows a kernel-smoothed histogram on each axis that illustrates the 1-dimensional distribution of the data.}
\end{center}
\end{figure*}

\begin{figure*}[!t]
\begin{center}
\includegraphics[width=0.8\linewidth]{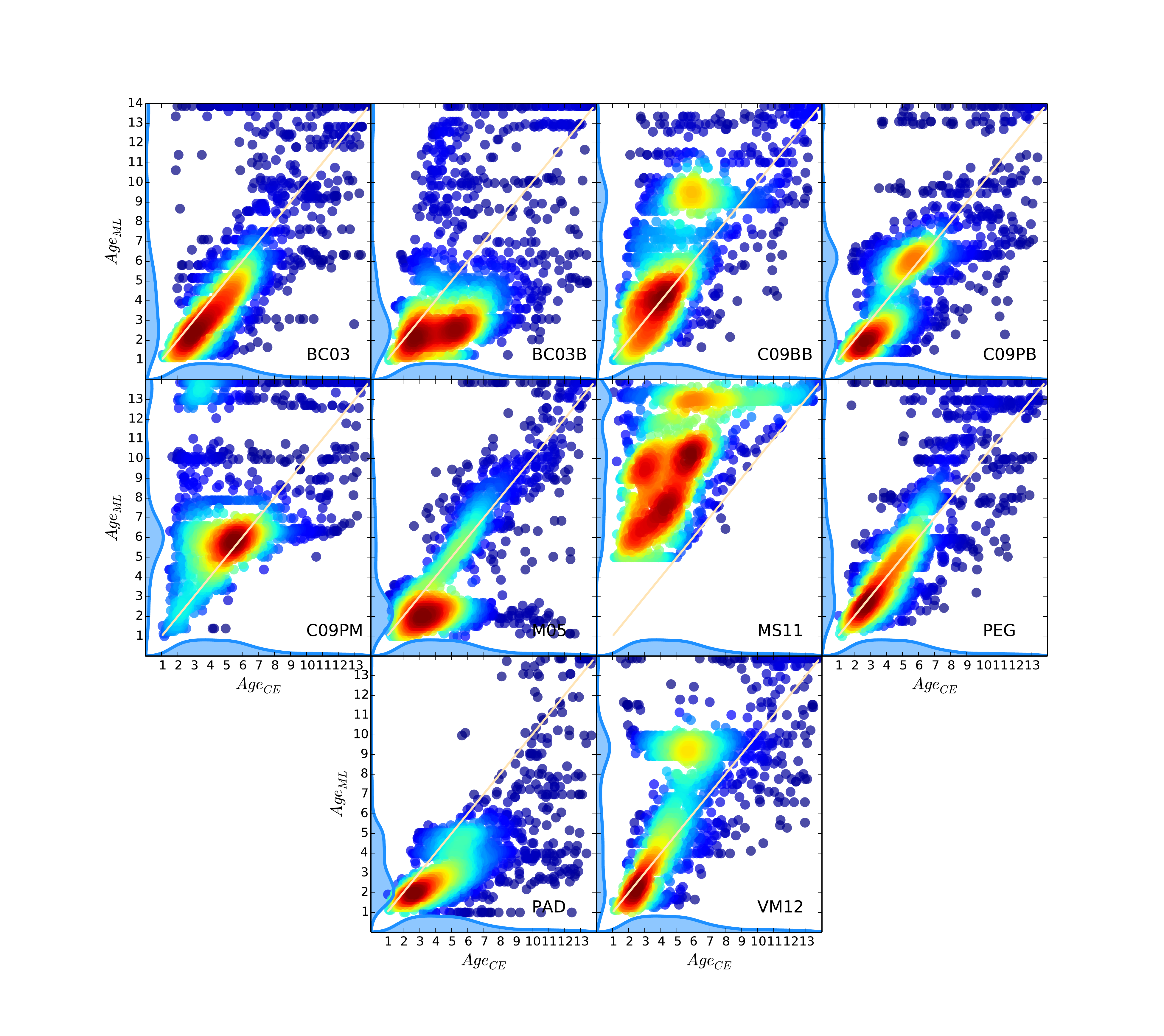}
\caption{\label{fig:ageage} Age distributions predicted by each SSP model (y-axis) compared with the CE (x-axis) for our entire Virgo core GC sample. The color shading of the symbols indicates the local data density, with dark red showing the highest concentrations.
 We show in beige the one-to-one line which highlights if a model is consistent with the CE i.e with a weighted mean of the other models. Each panel shows a kernel-smoothed histogram on each axis that illustrates the 1-dimensional distribution of the data.}
\end{center}
\end{figure*}

In an effort to understand how the age and metallicity estimates are affected by the model choice for the 1846 bright Virgo core GCs in our sample,
we compute the maximum-likelihood parameters with each SSP model individually.
Figure\,\ref{VirgoGCbanana} illustrates the age-metallicity likelihood distributions for a few examples, 
sorted by age horizontally and by metallicity vertically. 

In general, we find that for any given Virgo GC, the model-dependence of the age and metallicity estimates is much more significant than for a typical Galactic GC.
The detailed aspect of the figure depends on the cluster under study, but a generic feature is the frequent lack of overlap between the contours obtained with various models.
~Unfortunately, the differences between models, together with their inability to match the multi-wavelength locus of the data, erase the benefits of high quality photometry in passbands that should in principle allow us to better discriminate between age and metallicity effects.

In Figures\,\ref{fig:metmet} and \ref{fig:ageage}, we compare the age and metallicity distributions obtained
with each individual SSP model (y-axis) with the CE (x-axis).
~Using one particular model as a reference, instead of the CE, would have been justified if that model had been able to represent the color locus of Virgo GCs as a whole in a statistically satisfactory fashion.
But it was shown in Paper~I that none of the ten models satisfies this criterion.

~In the case of the metallicity distribution (Figure\,\ref{fig:metmet}), we observe good general agreement on the average values between the models.
However, individual models show particular deviations from the CE in some metallicity ranges, as illustrated by the curved density distributions (e.g. BC03B, C09PB, and C09PM).
Accumulations of data in the upper right corner of the diagram (i.e. the metal-rich end) and accumulations of data at one numerical value (e.g. in the PAD grid) occur when the  GC colors are beyond the borders of the age-metallicity
grid for that specific SSP model.~It is clear that the choice of a particular model can define the outcome of the metallicity distribution function.

For the inferred GC ages (Figure\,\ref{fig:ageage}), we observe significantly more outliers from 
the one-to-one relation than in the metallicity plots. 
For some models we find double-peaked distributions.
Accumulations of data at attracting values of age are common.  Again, these
occur mostly when the empirical colors lie outside the envelope of the grid of model colors; 
in some cases they are due to a non-monotonic evolution of model color with age, or 
to slightly sub-optimal sampling.

The individual estimates based on the ten models we have considered in this study, 
back up the results of the concordance estimate, i.e. a distribution with
a seemingly large fraction of young ages (and correspondingly relatively high metallicities).
Hence the concordance result is not the result of an inadequate weighting scheme 
between the models.
Taking the numbers at face value, 54\,\% of the GCs in our Virgo sample would have ages younger than 5\,Gyr,
and only about 10\,\% of the sample would be older than 9\,Gyr.
In Sections \ref{sec:ZPerrors} to \ref{sec:nirphot}, we quantify how various parameters 
may alter these distributions and in particular reduce (or not) the fraction of clusters that get 
assigned young ages. The studies in those sections all consider the same set of SSP models.

\subsubsection{Plausible Effects of Photometric Zero-Point Errors}
\label{sec:ZPerrors}

The age and metallicity differences we find between bright Virgo core GCs and Milky Way GCs are driven by a different location in color-color planes
that involve the ($u-g$), ($g-r$) or ($i-z$) colors \citep{powalka2016L}.
A similar displacement has been found recently by \cite{bellini2015}, based on {\it Hubble Space Telescope} (HST) observations of Milky Way GCs and of GCs in the central 3\arcmin\ of M87.
~These authors detect an offset of 0.06 magnitudes in the typical (F606W-F814W) colors between the two GC samples, at a given (F275W-F814W) color. The
similarity between their finding and ours is remarkable, considering their
data and ours are entirely independent.

\citet{bellini2015} note that the different measurement methods,
due among others to the different angular sizes of the objects on the sky,
could lead to systematic differences between the photometry of their Virgo
GCs and their MW objects,
the amplitude of which might erase some of the offset they see.
~Similarly, our photometry may suffer from systematic errors.
We have determined upper bounds on these for NGVS GCs in Paper~I
and will use these limits here, to evaluate to what extent they may
modify inferred age and metallicity distributions.

The error budget given in Table\,5 of Paper~I lists the estimated maximum systematic errors of our photometric measurements.
~We interpret these as 3\,$\sigma$ uncertainties, and start by exploring the effect of 1\,$\sigma$ zero-point errors in any direction in multi-dimensional color-color space, on GC ages and metallicities.
We then compare the fraction of GCs older than 9\,Gyr obtained for each of these directions, and isolate the vector for which the fraction is maximal.
The direction of this vector is shown in the top panels of Figure\,\ref{VirgoGCcolage}. As
expected, it also moves the colors of the Virgo core GCs closer to those of
the MW GCs. We note that selecting this direction is equivalent to suggesting that
the systematic errors had conspired to separate the samples from each other,
both in our data and in those of \citet{bellini2015}.

\begin{figure*}[!t]
\begin{center}
\includegraphics[width=8cm]{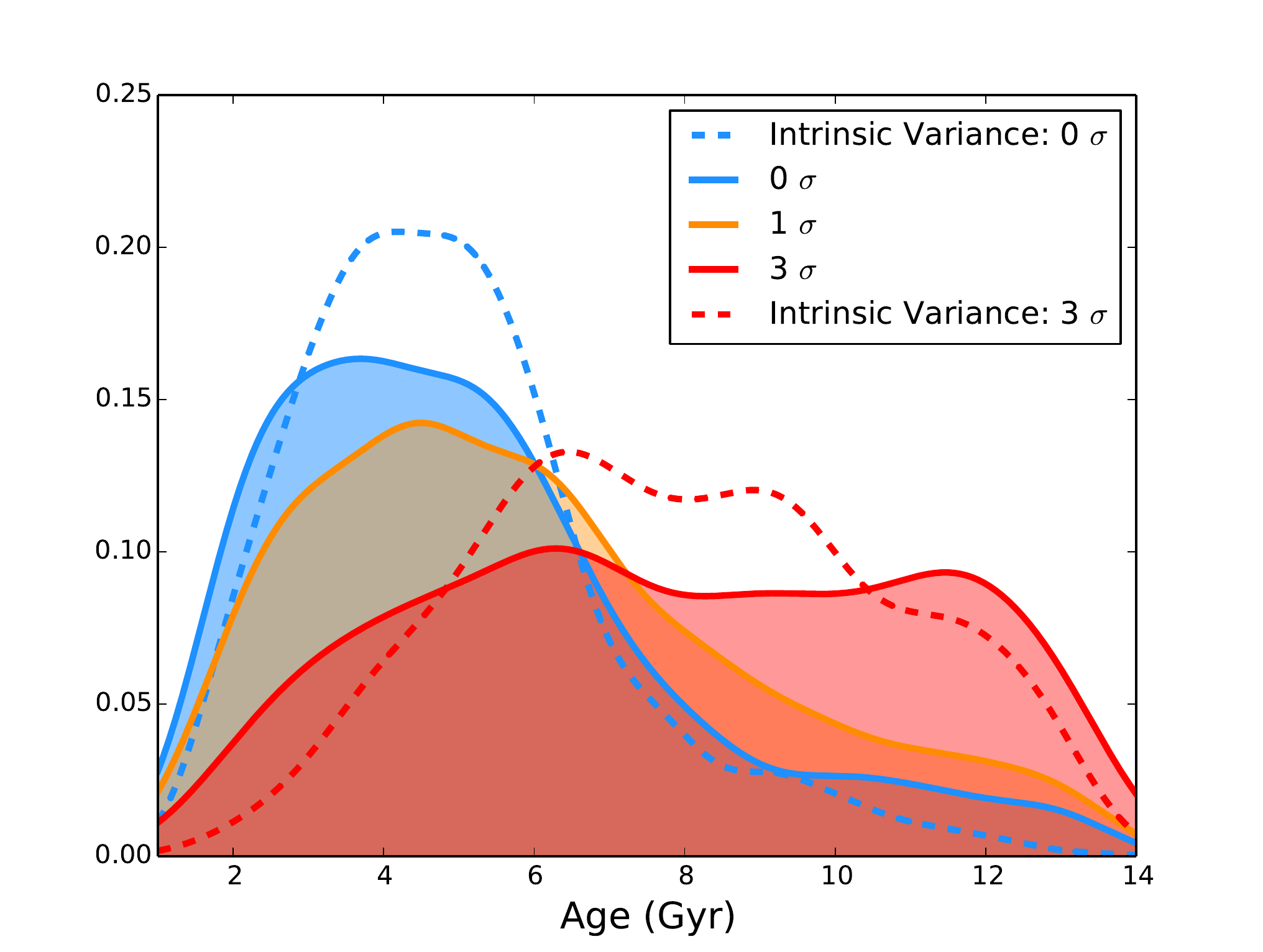}
\includegraphics[width=8cm]{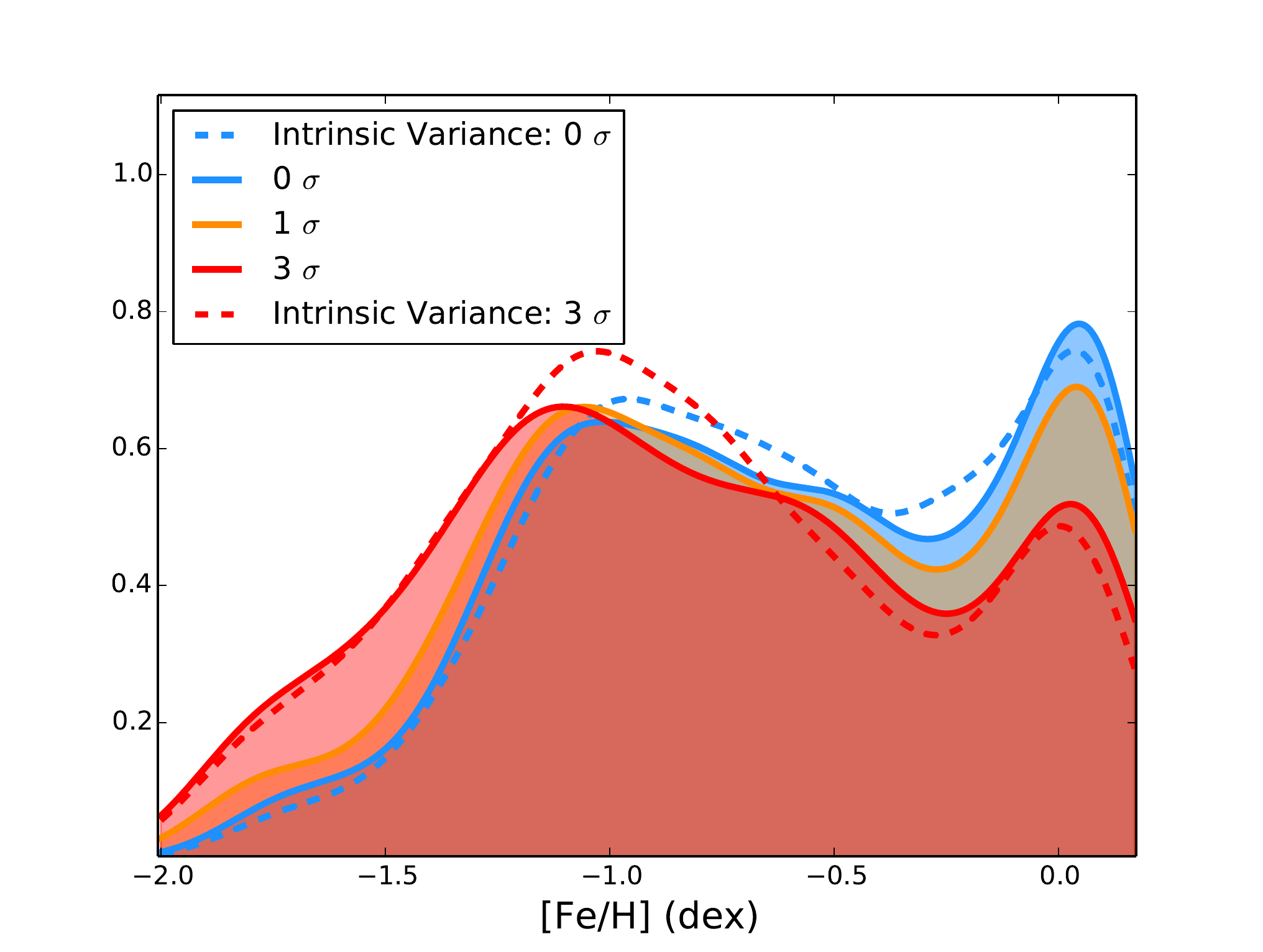}
\caption{\label{fig:offsethist} Photometric GC age and metallicity distribution functions (based on the 10-models CE) illustrated as kernel-smoothed distributions for 0\,$\sigma$ (blue), 1\,$\sigma$ (orange), 3\,$\sigma$ (red) offsets in the direction which creates the maximum fraction of old GC ages (see Fig.~\ref{VirgoGCcolage}). In addition, the blue and red dotted curves are, respectively, showing the 0\,$\sigma$ and 3\,$\sigma$ age distributions for a synthetic GC sample with colors based on the intrinsic variance described in Table\,\ref{tab:intvar} and defined in Section\,\ref{random_err}.}
\end{center}
\end{figure*}

\begin{figure*}[!t]
\begin{center}
\includegraphics[width=8cm]{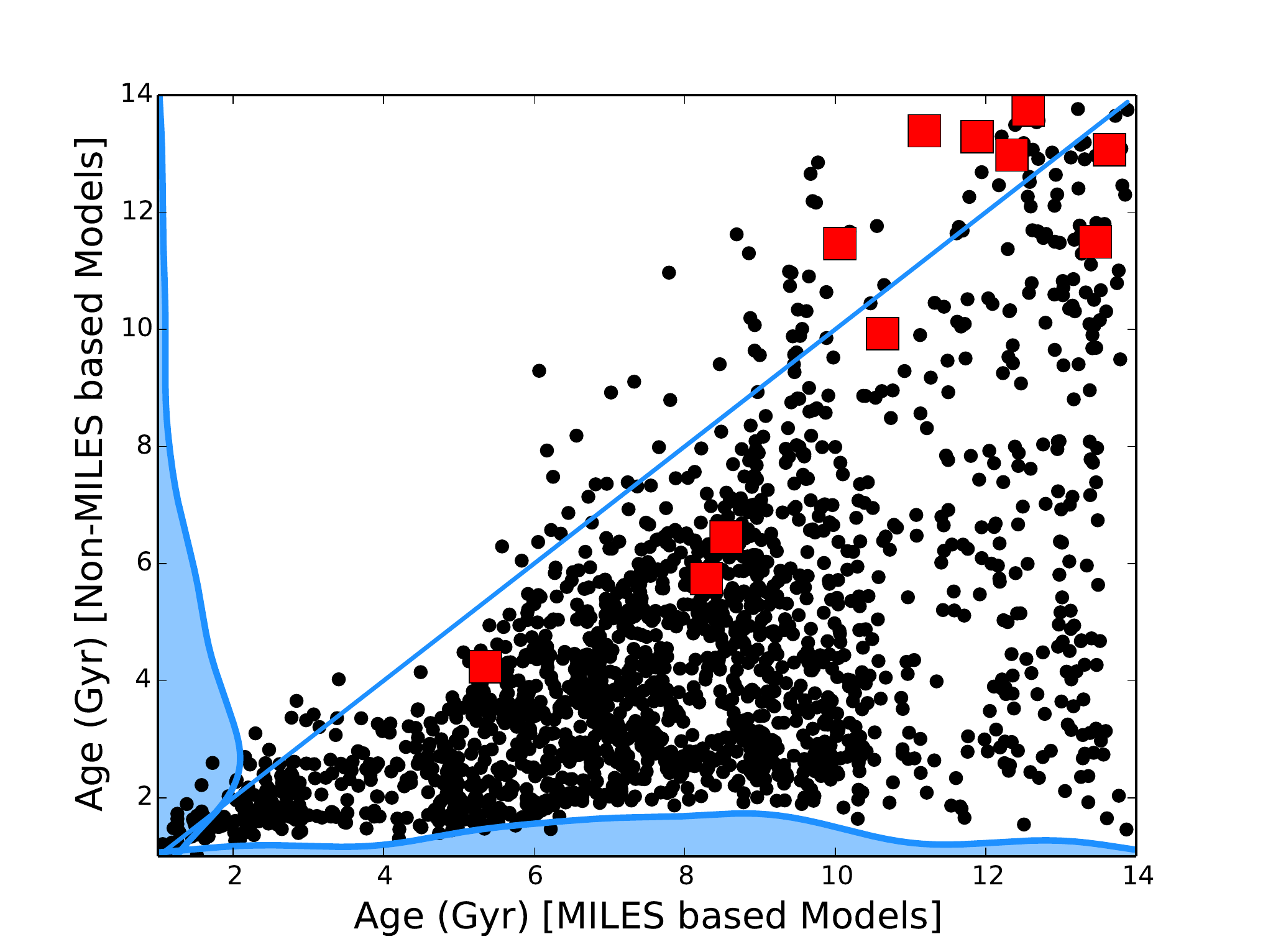}
\includegraphics[width=8cm]{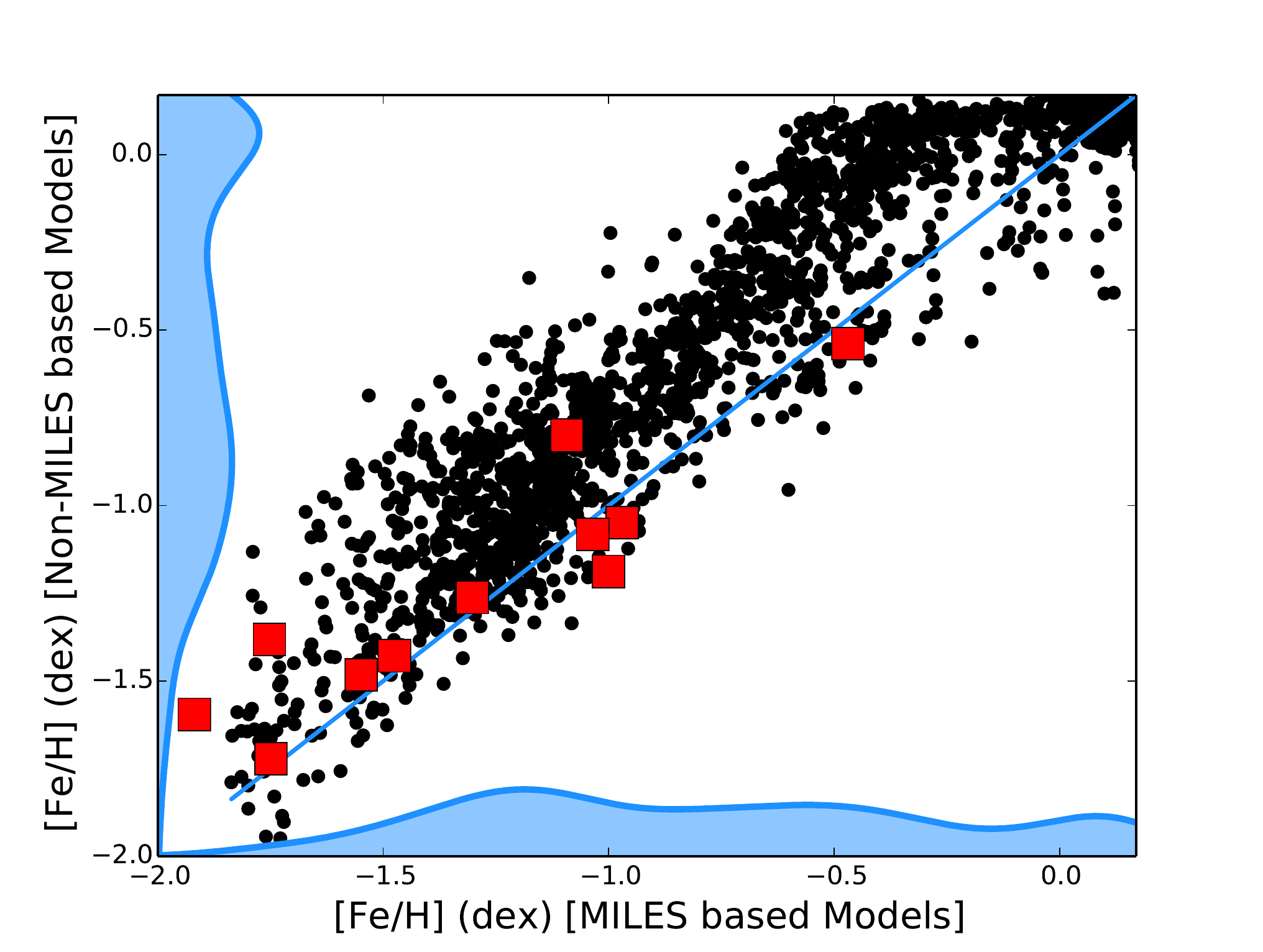}
\caption{\label{fig:lib_inf} Comparison of the age (left) and metallicity (right) distribution with different sets of model. On the x-axis, the CE is derived only using MILES based models i.e C09PM, MS11, and VM12, whereas
 on the y-axis, the CE is derived using the Non-MILES models i.e BC03, BC03B, C09BB, C09PB, M05, PAD, and PEG.
The Virgo GC sample is plotted with black dots and the MW GCs is showing with red square. The blue line is the one-to-one relation. Each panel shows a kernel-smoothed histogram on each axis that illustrates the 1-dimensional distribution of the data.}
\end{center}
\end{figure*}

\begin{figure*}[!t]
\begin{center}
\includegraphics[width=8cm]{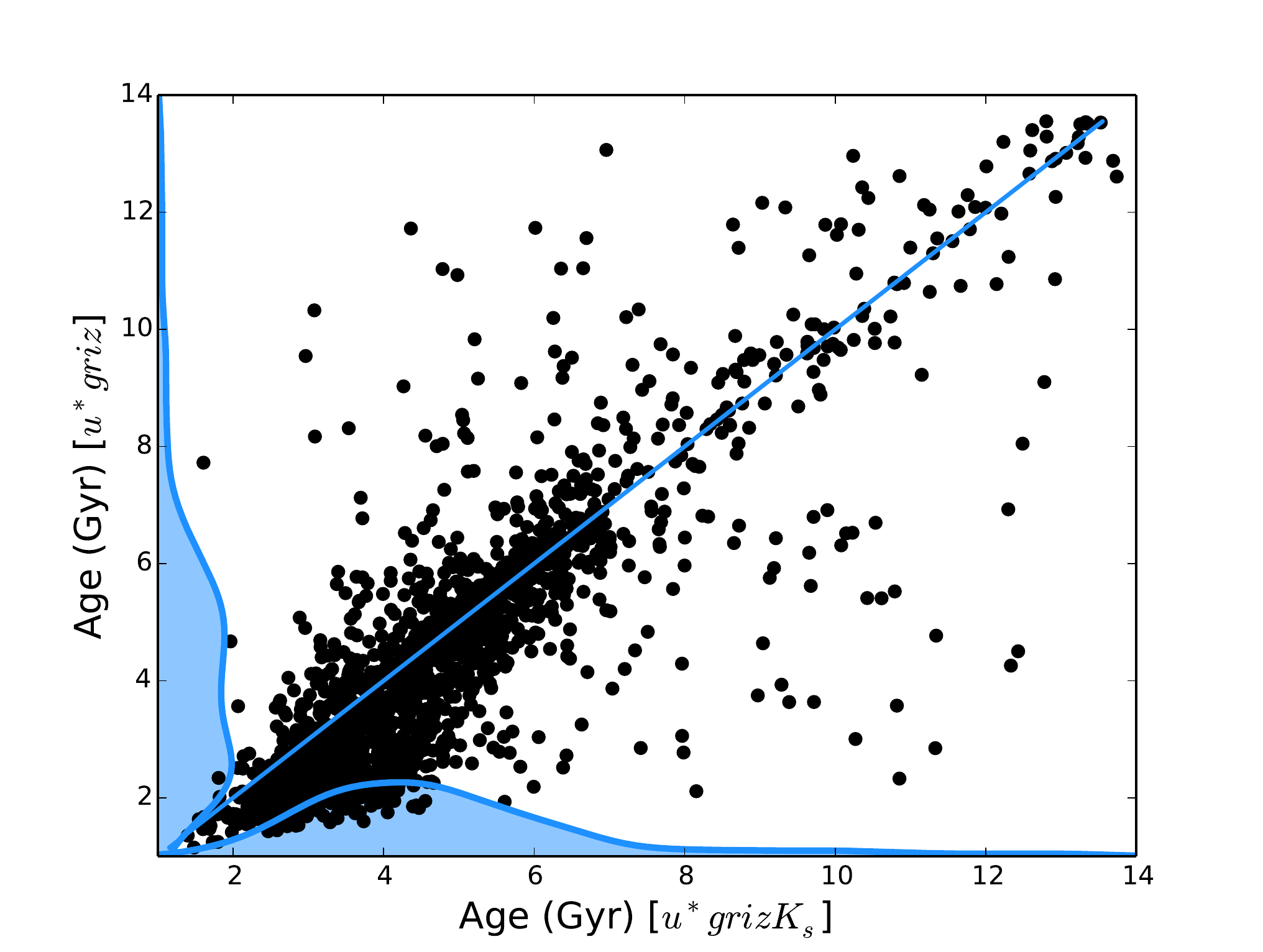}
\includegraphics[width=8cm]{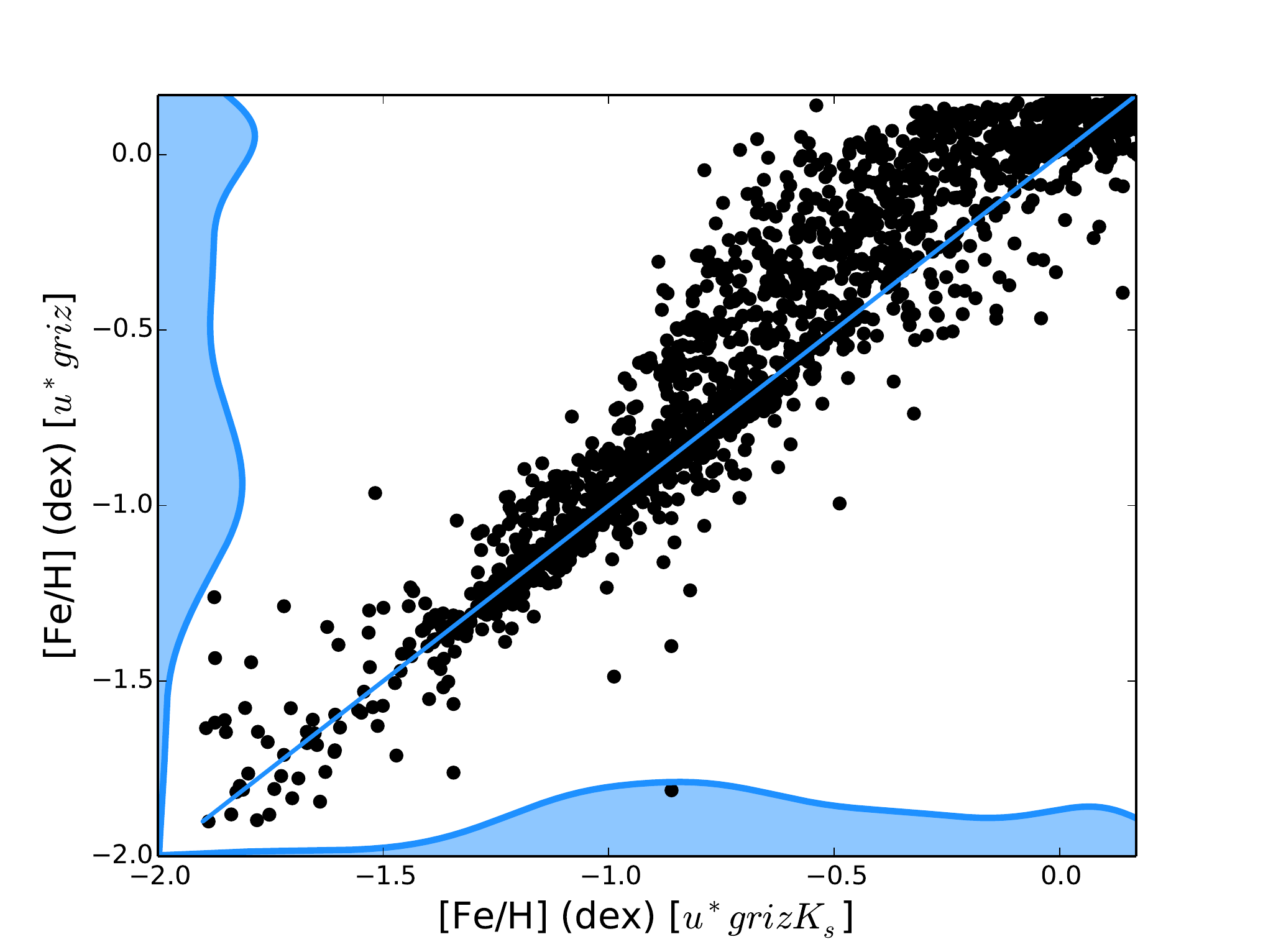}
\caption{\label{fig:ugriz_vs_ugrizk}  Comparison of the age (left) and metallicity (right) distribution with or without the $K_s$ band.
On the x-axis, we plot the 7-models CE with five optical/NIR colors ($u-g$, $g-r$, $r-i$, $i-z$, and $g-K_s$) whereas on the y-axis we derive the 7-models CE with four optical colors ($u-g$, $g-r$, $r-i$, and $i-z$).
The blue line is the one-to-one relation. Each panel shows a kernel-smoothed histogram on each axis that illustrates the 1-dimensional distribution of the data.}
\end{center}
\end{figure*}

The age and metallicity distributions obtained for our sample of bright Virgo 
core GCs, for shifts of 1 and 3\,$\sigma$ 
along the selected direction (see vectors in the top panels of Fig.~\ref{VirgoGCcolage}), are shown as orange and red distributions in Figure\,\ref{fig:offsethist}. 
The average age returned by the CE analysis of the offset colors (3\,$\sigma$) is about 3\,Gyr 
older than with the original data.  Correspondingly, fewer GCs are 
assigned high metallicities. The average metallicity however changes only by
about 0.20\,dex.

Before returning to the fraction of sample GCs with assigned young ages, we
consider the broadening effect of random errors on the distribution functions.

\subsubsection{Broadening by Random Photometric Errors}
\label{random_err}
The photometric uncertainties on the observed GC colors are small
in our data set, but they are responsible for a significant fraction of the
variance across the main locus of the GC observations in color-color space.
In Paper~I, we attempted to represent the GC distribution with a 1-dimensional
fiducial locus, and concluded that the reduced $\chi^2$-distance of the
GCs to that locus was too large for this 1-D locus to be an acceptable parent
distribution of the data. Some of the dispersion is real.
Looking at this question in more detail, we can identify regions of the color locus 
in which random errors explain the observed variance, and regions in which they do not. 
This impacts the interpretation of the width of the inferred age and metallicity
distributions.

We investigate the effects of random errors with simulated cluster samples.
1846 artificial GCs are placed on a fiducial 1-dimensional polynomial locus
in color space, that represents the typical locus of GCs near M87 and
is obtained as described in Paper~I.
Their ($g-K_s$) values are those of the observed
1846 Virgo GCs, and their other colors are as given by the 1-D locus.
To the magnitudes of the artificial GCs, we add errors drawn from
normal distributions with standard deviations equal to those of the
corresponding empirical clusters. The artificial object colors
are then analysed for age and metallicity using the CE method.

\begin{figure*}
\begin{center}
\includegraphics[width=0.45\textwidth]{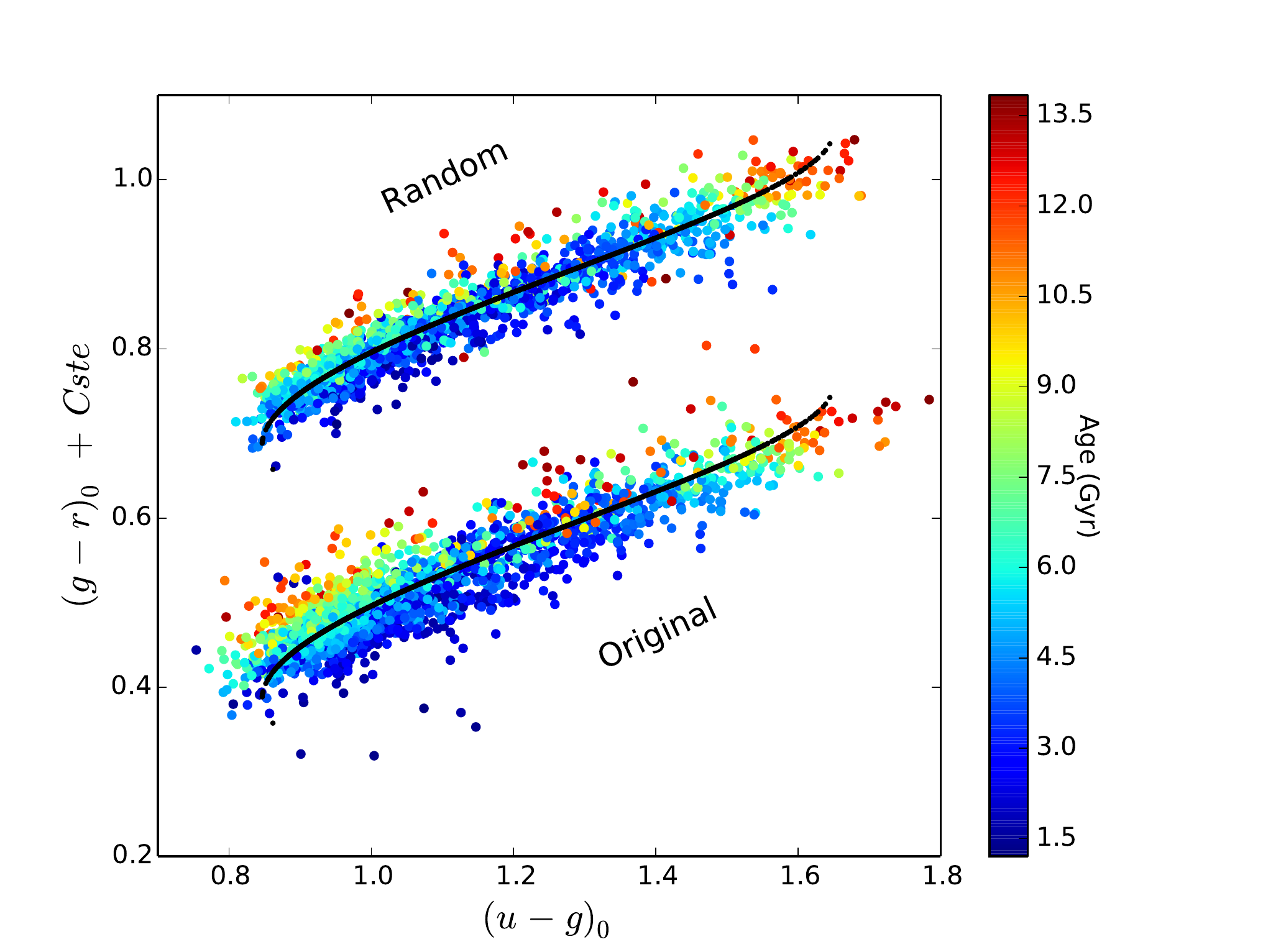}
\includegraphics[width=0.45\textwidth]{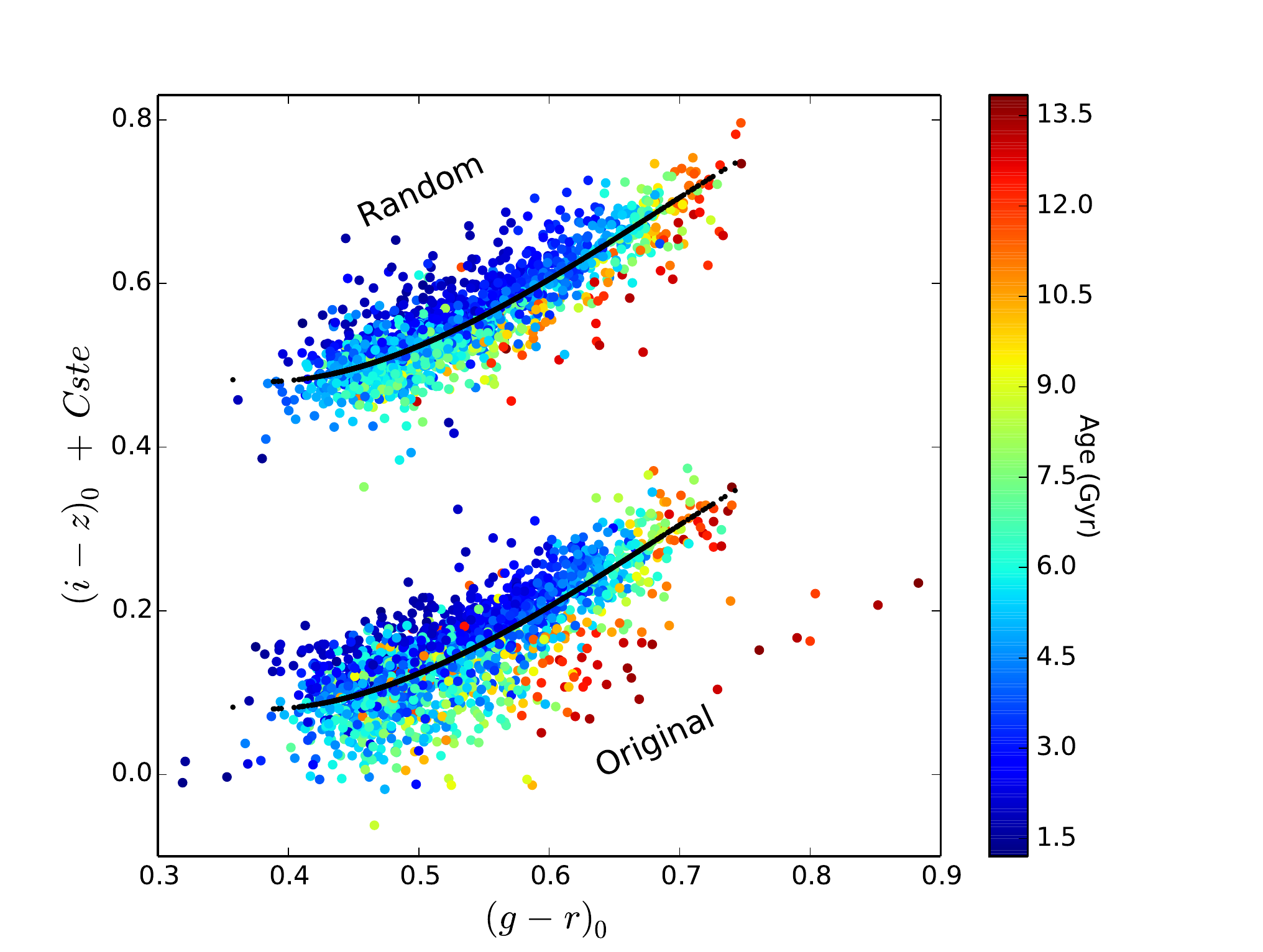}
\caption[]{Observed color distribution of our sample of Virgo GCs, and artificial distribution
obtained assuming an infinitely narrow parent distribution and random errors distributed as in
the observations (offset vertically for clarity).}
\label{fig:RandomVsObs}
\end{center}
\end{figure*}

The broadening of the GC locus due to random errors is comparable
to the observed broadening at the red end of the distribution, but is narrower
elsewhere (Figure\,\ref{fig:RandomVsObs}).
In each color-color diagram, we quantify the intrinsic variance
as a function of ($g-K_s$) as the difference between the variance 
observed orthogonally to the 1-D fiducial locus, and the variance obtained
in the same direction in the artificial
distribution. The results are given in Tab.\,\ref{tab:intvar}.
The first column describes position along the color locus.
Columns (2) to (5), (6) to (9) and (10) to (13), respectively,
provide the observed dispersion (as a standard deviation), 
the dispersion in the artifical
sample, and the resulting intrinsic dispersion of the colors of
the observed sample.

\begin{deluxetable*}{lcccccccccccc}[b]
\tablecaption{\label{tab:intvar} Bright Virgo core GCs: observed and intrinsic
color spread across the main locus.}
\tablehead{\colhead{$g-Ks$} & \colhead{$u-g_{\rm ori}$} & \colhead{$g-r_{\rm ori}$} & \colhead{$r-i_{\rm ori}$} & \colhead{$i-z_{\rm ori}$} & \colhead{$u-g_{\rm ran}$} & \colhead{$g-r_{\rm ran}$} & \colhead{$r-i_{\rm ran}$} & \colhead{$i-z_{\rm ran}$} & \colhead{$u-g_{\rm iv}$} & \colhead{$g-r_{\rm iv}$} & \colhead{$r-i_{\rm iv}$} & \colhead{$i-z_{\rm iv}$}\\
(1) & (2) & (3) & (4) & (5) & (6) & (7) & (8) & (9) & (10) & (11) & (12) & (13) \\
}
\startdata
  0.310 & 0.045 & 0.030 & 0.022 & 0.055 & 0.022 & 0.021 & 0.013 & 0.024 & 0.039 & 0.020 & 0.018 & 0.050\\
  0.521 & 0.045 & 0.024 & 0.021 & 0.047 & 0.026 & 0.019 & 0.014 & 0.026 & 0.036 & 0.012 & 0.016 & 0.038\\
  0.733 & 0.047 & 0.025 & 0.021 & 0.040 & 0.030 & 0.020 & 0.019 & 0.029 & 0.036 & 0.014 & 0.008 & 0.028\\
  0.945 & 0.050 & 0.026 & 0.023 & 0.037 & 0.036 & 0.022 & 0.020 & 0.030 & 0.033 & 0.013 & 0.011 & 0.022\\
  1.157 & 0.053 & 0.024 & 0.022 & 0.035 & 0.041 & 0.021 & 0.021 & 0.030 & 0.033 & 0.010 & 0.006 & 0.018\\
  1.369 & 0.059 & 0.032 & 0.021 & 0.035 & 0.050 & 0.021 & 0.020 & 0.031 & 0.030 & 0.023 & 0.004 & 0.017\\
  1.581 & 0.055 & 0.029 & 0.021 & 0.036 & 0.059 & 0.017 & 0.021 & 0.026 & x & 0.022 & 0.004 & 0.024\\
  1.793 & 0.054 & 0.027 & 0.022 & 0.037 & 0.061 & 0.019 & 0.017 & 0.029 & x & 0.019 & 0.014 & 0.023\\
\enddata
\tablecomments{The values are standard deviations in magnitudes.}
\end{deluxetable*}

Using the results in Tab.\,\ref{tab:intvar}, we may re-evaluate the 
width of the CE-based age and metallicity distributions after correction
for the broadening by random errors. To this purpose,
we construct a sample of synthetic GCs whose intrinsic
colors are distributed around the fiducial locus according to the
intrinsic variance described in Tab.\,\ref{tab:intvar}. The results of their analysis 
are overlaid in Figure\,\ref{fig:offsethist} as dotted kernel-smoothed histograms.
The resulting ranges of both synthetic age and metallicity distributions are comparable with those of the original distributions.
It means that the sole random errors are not artificially creating our age distribution. 
Even when accounting for random errors and when considering unfavourable systematic errors, the analysis 
with standard SSP models inevitably leads to a distribution with a seemingly large 
fraction of young ages.

\subsubsection{Influence of the stellar spectral library}
\label{inf_lib}

The conclusion of Paper~I emphasized that the stellar spectral libraries used in the EPS models have a strong influence on the predictions in color-color space.
In particular, we saw that the models based on the MILES library tend to reproduce the SED around the $g$ band better, whereas the ones using the BaSeL library better match the SED of bright Virgo core GCs in the $z$ band (the $z$ band lies outside of the MILES wavelength range, hence MILES must be patched with other libraries for these predictions; see Tab.\,\ref{tab:ssp}).  

In this Section, we assess the inferred age-metallicity difference related to the influence of the library. 
To this end, we derive metallicity and age using two different sets of SSP models.
For the first one, we use only models with the MILES library i.e C09PM, MS11, and VM12. For the second one, we take all the models based on other libraries than MILES i.e BC03, BC03B, PEG, PAD, M05, C09BB, and C09PB.
We show in Figure\,\ref{fig:lib_inf} the age and metallicity obtained with both SSP samples along with the associated histograms for the Paper~I Virgo sample (black dots) and for the MW GCs (red squares).
For the NGVS GCs, we observe that the estimates with the MILES library tend to favour older ages and lower metallicities. The age histogram of the sample 
is composed of three peaks, one main between 5\,Gyr and 9\,Gyr and
two others at 13\,Gyr and 3\,Gyr. The old one (age$\,>\,9$\,Gyr) contains 33\,\% of the GC sample whereas the intermediate age one (age$\,>\,5$ and $\,<\,9$\,Gyr) contains 51\,\% and the young one (age$\,<\,5$\,Gyr) 16\,\%.
On the contrary, the estimates based on the other libraries tend to produce younger ages with a main peak at 3\,Gyr and higher metallicities. This histogram gives 67\,\% of young GCs (age$\,<\,5$\,Gyr), 25\,\% of intermediate age one,
and 8\,\% of old GCs.
Interestingly, for the MW GCs, we do not observe any systematic age and metallicity differences.
But for one model set more than half the NGVS Virgo core GCs are assigned ages older than 8\,Gyr, 
whereas for the second set half of the NGVS GCs are given ages below 4\,Gyr.
This extreme age difference mainly arises from the color difference
observed between the MILES and the BaSeL, STELIB, ATLAS ODFNEW / PHOENIX BT-Settl libraries, 
particularly in the $g$ band. We note that the MS11 model grid does not include ages 
below 5\,Gyr which could in principle bias the estimates toward older ages, but Figure\,\ref{fig:ageage} shows 
no pile-up of inferred ages at the lower age limit of the grid, suggesting this bias is not severe\footnote{In
contrast, running the analysis with other model grids after truncating these at 5\,Gyr does produce 
a pile-up.}.

It is worth mentionning that the ages our procedure returns with the MILES set of models are more consistent 
with prior expectations than the younger ages inferred with all models. In the MILES-based distribution, the old 
GCs could be formed in situ in the primordial M87 galaxy whereas the intermediate GCs could be created 
during the multiple mergers on M87. Because derived distributions still depend on the colors used, and 
because the MILES-based models are not a fully satisfactory representation of all the SEDs, we do 
not suggest only a subset of models should be kept and the others discarded.  We simply 
emphasize with this study the strength of the model dependence of the results.

\subsubsection{Near-IR photometry}
\label{sec:nirphot}

Past studies have suggested that the inclusion 
of near-infrared passbands in the analysis of SEDs helps break the
age-metallicity degeneracy \citep[e.g.][]{puzia2002, kisslerpatig2002,
montes2014}. The extent 
to which this is expected to be true depends on the exact color combination used
\citep{anders2004}, and also on the particular population synthesis
code adopted (see Figures 12 to 16 of Paper~I for examples of synthetic color
grids parametrized by age and metallicity).
$K_s$ magnitudes from NGVS-IR are available for the 1846 bright GCs 
of our Virgo core sample. Here, we briefly examine if added near-IR information
modifies the age and metallicity distributions inferred with our CE method.

Unfortunately, we cannot compute $u^*grizK_s$ colors with every population 
synthesis code, because some do not reach far enough into the near-IR.
Therefore, we compare the age-metallicity estimates derived with and without the $K_s$ band using only seven models: BC03, BC03B, C09PB, C09PM, M05, PEG, PAD.  
We show in Figure\,\ref{fig:ugriz_vs_ugrizk} both age and metallicity estimates computed with ($u-g$), ($g-r$), ($r-i$), ($i-z$), ($g-K_s$) colors (x-axis) or only with ($u-g$), ($g-r$), ($r-i$), ($i-z$) colors (y-axis).

It appears that the inclusion of the $K_s$ photometry does not produce systematically older ages.
The metallicity is also only slightly affected, with a just slightly larger fraction of metal-poor GCs 
(near [Fe/H]\,$\sim$\,-0.5) when $K_s$ is included.
For the two sets of colors used, the influence of the $K_s$ band is weak. It is likely that the age-metallicity 
constraints imposed by the $K_s$ band are diluted with the 4 other optical colors used in combination
with it, and that the expected benefits of adding $K_s$ are partly lost because the models do not reproduce the colors as a whole well enough.

\section{Discussion}
\label{sec:discussion}

In Section\,\ref{sec:results}, we have shown that the analysis of the
$u^*griz(K_s)$ colors of bright Virgo core GCs, with a collection of standard 
SSP models, generically leads to distributions of assigned ages and metallicities 
with strong extensions into the regime of young ages, the seemingly young
clusters also having relatively high assigned metallicities. This is 
the result of the location of these GCs in color-color space, which
differs from the locus of the GCs in the Milky Way. On average, the
predictions of standard population synthesis models are closer to the 
locus of the MW globular clusters, and the bright Virgo core GCs lie
on the young side of the synthetic SSP color-distributions.

Formal age distributions with tails towards intermediate or young ages 
have been found previously by various authors, for GCs in a variety of environments. 
For instance, \citet{puzia2005a} found several M31 GCs 
with estimated ages around 5\,Gyr and 8\,Gyr, based on optical spectrophotometric
indices.
\citet{park2012} assessed several giant elliptical galaxies,
again using spectrophotometric indices,
and concluded the age range of their GCs was broad (from 1\,Gyr to 14\,Gyr). 
\citet{usher2015} combined ($g-i$) colors with stacked optical and calcium-triplet 
spectra of GCs around various elliptical galaxies, and found ages below 5\,Gyr 
in a small fraction of their galaxy-and-color bins.
Based on HST and VLT (adaptive optics) photometry 
across the ultraviolet, visible and near-IR,
\citet{montes2014} found that 20\,\% of the innermost GCs of M87
seemed to have young ages.

In comparison with those results, the fraction of young ages obtained 
for Virgo core GCs when using $u^*griz(K_s)$ photometry and ten standard SSP models, 
appears extreme. For this particular passband combination, only a few individual SSP model grids
assign the clusters mostly intermediate and old ages (some of those based on MILES stellar spectra).  
While the young ages of the concordance estimates are difficult to reconcile with the current 
views on GC formation, the results for these particular models might be more plausible. 
Although we caution that this may be incidental (for even these models 
account for the expected complexity we come back to later in this section),
we briefly examine paths that could lead to relatively young objects. 

One way of forming clusters late is to invoke mergers of gas-rich galaxies
\citep{whitmore1993,whitmore1995,renaud2015}. 
Morphological signatures of major mergers disappear on timescales of a few Gyr
\citep[][$\sim$\,2-3\,Gyr;]{borne1991}.
Unfortunately no detailed high resolution simulations are available to test
the formation and survival of globular clusters in a major merger that would occur in
as dense and hot an environment as the Virgo cluster.
In a recent study, \citet{ferrarese2016} used the low mass luminosity function of  
galaxies in the Virgo core
to estimate the number of galaxies that might have disrupted over time in what is now 
M87. The resulting numbers, though uncertain, are of the order of $\sim 10^3$ galaxies. 
In this scenario, these galaxies altogether could have brought in as many as 40\,\% of 
the current GC population of M87. In order to produce a large proportion of GCs 
with intermediate ages this way, one should expect several large wet mergers 
to have occured between incoming galaxies before their disruption. This sets rather 
fine-tuned timing constraints, and again simulations would be needed to test this further. 
Finally, a fraction of the massive star clusters in our Virgo core sample may in
fact consist of previous nuclei of infalling galaxies, of which some could contain
young or intermediate age populations \citep{georgievboeker14}. 

\smallskip

More likely, the photometric ages derived with standard SSP models may simply 
be the result of using unsuitable assumptions for the synthetic stellar populations.
Standard SSP models were initially built to reproduce Galactic stellar populations,
and indeed we find they lead to results globally consistent with detailed, resolved
studies, for a set of Milky Way GCs. Considering the color differences observed between the 
MW and the M87 GCs, a single relation between colors and the fundamental parameters
age and metallicity is essentially excluded \citep{powalka2016L, usher2015}. 
Our study emphasizes the need for population 
synthesis models specifically tuned to the particular stellar population properties of the Virgo 
environment.

The M87 GCs are likely not ideal single stellar populations. 
In the Milky Way, the color-magnitude diagrams of massive star clusters 
exhibit multiple stellar populations, i.e. neither the GC abundances nor their ages 
should be described with a unique value \citep{piotto2015, renzini2015}. 
The Virgo clusters studied here are all massive,
some of them may be previous galaxy nuclei, hence they are likely to be at least
as complex as MW GCs.

As mentioned in Section\,1, it is known that higher-than-standard proportions 
of old but blue stars affect colors and spectra in a way that mimics young ages 
\citep[][]{cenarro2008,koleva2008,xin2011,chiessantos2011}.
\citet{lee1994} summarized the wide variety of HB morphologies in the MW GCs. 
Blue horizontal branches may be due to peculiar helium abundances \citep[e.g.][and
references therein]{renzini2015}, to strong winds on the red giant branch \citep[e.g.][and
references therein]{buzzoni2008} or to tidally
enhanced stellar winds in binary systems \citep{lei2015}.
As a first step, some SSP modelling codes include an option for a red or a blue HB morphology. 
Using M05/M11 \citep{maraston2005,maraston2011}, we found that the red-HB version was a better representation of the
locus of Virgo core GCs in color-color space than the blue-HB version.
Blue stragglers or hot subdwarfs are other possible old contributors to the short
wavelength light \citep{han2007}. \citet{fan2012} compared models with and without
blue stragglers and found the effect on estimated ages could reach 0.3\,dex. For a 10\,Gyr old
GC, this translates into 5\,Gyr.
Photometry at shorter UV-wavelengths than used here could in principle help 
evaluate the effective temperatures of the hottest stars present. An attempt to identify the 
contribution of blue horizontal branches using HST imaging of M87 GCs at 275\,nm concluded it 
was too small to be measured, except for a small subpopulation of extreme objects
\citep{bellini2015}.

More generally, no non-solar abundance ratios and no internal chemical abundance spreads 
are taken into account in the standard SSP models we have used. For instance, the influence 
of the helium content in stellar isochrones \citep{chantereau2015},
the CNO abundance variation \citep{aringer2016}, or the possible $\alpha$-element enhancement 
\citep{lee2009,vazdekis2015} are known to modify the GC SED. 
In our recent letter \citep{powalka2016L},
we suggest that a combination of these effects could produce the color difference 
observed between the M87 and the MW GCs.
In particular, we briefly assessed the modification produced by an 
$\alpha$-enhancement of [$\alpha/$Fe]\,=\,0.4 in the input stellar spectral library of the models. The tests were based on the population synthesis code {\sc Pegase} coupled with 
synthetic stellar spectra from \citet{husser2013}. Although the direction of the shift may
explain part of the observed offset between these two GC populations,
its amplitude is 4 $\times$ lower than needed. 
In a very indirect way, the fact that current solar-scaled models provide a reasonable match to MW GCs, despite the known [$\alpha/$Fe]--[Fe/H] anti-correlation among these objects, also suggests $\alpha$-elements are not the only players.
More complex changes in the abundances patterns are probably present 
between Virgo and the Milky Way. Therefore, next generations of 
stellar population models will have to include stellar isochrones that account 
for a variety of initial abundance patterns as well as chemical variation along stellar evolution tracks \citep[e.g.][]{georgy2013,dotter2017},
and stellar libraries with spectra for different chemical abundance patterns.
Efforts in these directions are underway \citep{lee2009,coelho2014,vazdekis2015,aringer2016,choi2016}, but it will take some time to obtain models that combine 
all the relevant effects and extend from the near-UV to the infrared. As yet, we found no single model grid that reproduces
the observed locus of Virgo core GCs in $ugriz(K_s)$ color-space -- once this shortcoming 
is fixed, we may be able to extract more stringent constraints from photometric surveys.

\section{Summary and conclusion}
\label{conclusion}

This paper provides an update on photometric age and metallicity estimates for GCs with
high quality, multi-waveband integrated photometry, in different environments.

In previous articles of this series, we acquired 1) the $u^*grizK_s$ MegaCam/WirCam magnitudes for a sample of Virgo core GCs using the NGVS and 2) the $u^*griz$ photometry for a set of 11 MW GCs using well calibrated, integrated VLT/X-shooter spectra.
In this paper, we compared these two samples with ten commonly used SSP models using four colors ($u-g$, $g-r$, $r-i$, $i-z$), or five (with $g-K_s$).
We used a weighted scheme to defined a concordance estimate, called CE, which gives the photometric age and metallicity estimates on which the ten models tend to agree.

First, for the MW GCs, we verified the consistency of our estimates with those in the literature. Except for one MW GC (NGC\,288), we found similar results, i.e. old ages 
and mostly metal-poor compositions.

Then, we applied our CE to the Virgo core GCs. We found that the large majority of our 1846 GCs are assigned formal ages younger than 9\,Gyr with this method. Although, some
earlier studies have found intermediate age GCs (between 5 and 9\,Gyr) in galaxies, we raise serious doubts about the absolute ages the comparison of the Virgo GC colors with stan-
dard population synthesis models produces.
We investigated the possible origins of these results.

First, we used the budget of systematic errors available in Paper~I in order to check to what extent these errors could have
conspired to produce seemingly young ages. When applying maximal zero-point shifts in the direction in 5-dimensional
space that maximizes the fraction of old ages, we were only
able to raise the fraction of inferred old ages ($\,>\,$9\,Gyr) to 39\,\%.
Regarding the metallicity, we only observed a small influence of the shift on the results.

In addition, we investigated the age dispersion caused by the random errors of our sample. To do so, we constructed a artificial GC sample corrected from the broadening by the random errors.
The resulting ranges of both age and metallicity distributions still appeared to be comparable with the original sample.
We concluded that the random errors are not responsible for the wide range of photometric ages found for Virgo core GCs. 

Considering the conclusions of Paper~I about the influence of stellar spectral libraries on SSP model colors, we reassessed ages and metallicities, with models using the MILES
stellar library on one hand, and with models using other libraries on the other.
We found that the MILES-based set favored older Virgo GC ages, with 51\,\% intermediate ages (5-9\,Gyr) and 33\,\% old ages ($>\,$9\,Gyr). This last percentage can rise up to 67\,\% when, in addition to using only MILES-based
models, one corrects for maximal zero point errors in the direction of 5-D space producing maximal effect.
Conversely, the other set predicts only 25\,\% intermediate ages and 8\,\% old ages (with a maximum of $\sim\,$30\,\% with the maximal color shift).
Surprisingly, the estimates for the MW GCs did not show any systematics as a function of the model set.

Finally, we studied the influence of the $K_s$ band on the Virgo GC parameter estimates. Although several studies recommend the use of the $K_s$ band to break the age-metallicity degeneracy,
the addition of a fifth color ($g-K_s$) in our computation gave results similar 
to those with four colors. In the absence of a real modification, we concluded that the
$K_s$ band constraints might be diluted with the four other optical colors.

We emphasize that the proportion of young assigned ages in our standard-SSP based study of
the Virgo sample is related to the color difference and environmental effect 
that we highlighted in \citet{powalka2016L}.
In other words, the calibration of the current SSP models on MW GC data along with the color difference observed between the MW and M87 GCs necessarily induce different photometric age estimates for the two samples when the analysis is performed with the same sets of models.
The possible origins of these color differences, other than age and metallicity, are numerous (chemistry, rotation, binarity, extended star formation histories, dynamical effects affecting the stellar mass function, etc.).
Anticipating future papers, we note that the colors of several bright M49 GCs are similar to those of M87, inducing similar systematic offsets with respect to the MW. 

From an alternative point of view, a real GC age difference could be partially responsible for the observed color discrepancies between the GCs from the MW and M87.
Although the age difference is not expected to be as high as found in this study, the rich merger history of M87 (in constrast with the MW) lends credence to the formation of distinct stellar populations.

We conclude this paper by a caveat on the absolute age and metallicity estimates.
Despite the high quality of the photometric data used here (and expected in future surveys), we cannot as yet reasonably use photometric ages and metallicities to unravel the detailed assembly history of galaxies such
as M87.
Obtaining models that actually match the multidimensional color-locus of modern collections of GC observations is a formidable task, but it is becoming more and more
necessary.
The narrow locus of Virgo core GCs in color-color planes will provide strong constraints on new generations of models.

\acknowledgments

This project is supported by the FONDECYT Regular Project No.~1161817 and BASAL Center for Astrophysics and Associated Technologies (PFB-06).
We gratefully acknowledge the support by the French-Chilean Collaboration Program ECOS Sud-CONICYT under grant C15U02 as well as support from the French Programme National de Physique
Stellaire (PNPS, AO2016) and Programme National Cosmologie \& Galaxies
(PNCG, AO2017).
EWP acknowledges support from the National Natural Science Foundation of China through Grant No. 11573002, travel support from the Chinese Academy of Sciences South America Center for Astronomy, and the Sino-French LIA-Origin joint exchange program.
C.L. acknowledges the National Key Basic Research Program of China (2015CB857002) and the NSFC grants 11673017, 11203017, 11433002.  CL is supported by Key Laboratory for Particle Physics,Astrophysics and Cosmology, Ministry of Education, and Shanghai Key Laboratory for Particle Physics and Cosmology(SKLPPC).
S.M. acknowledges financial support from the Institut Universitaire de France (IUF), of which she is senior member.

\appendix
\section{Appendix A: Comparison with Cohen et al. (1998)}
\label{setvscohen}
\begin{figure*}
\begin{center}
\includegraphics[width=0.8\textwidth]{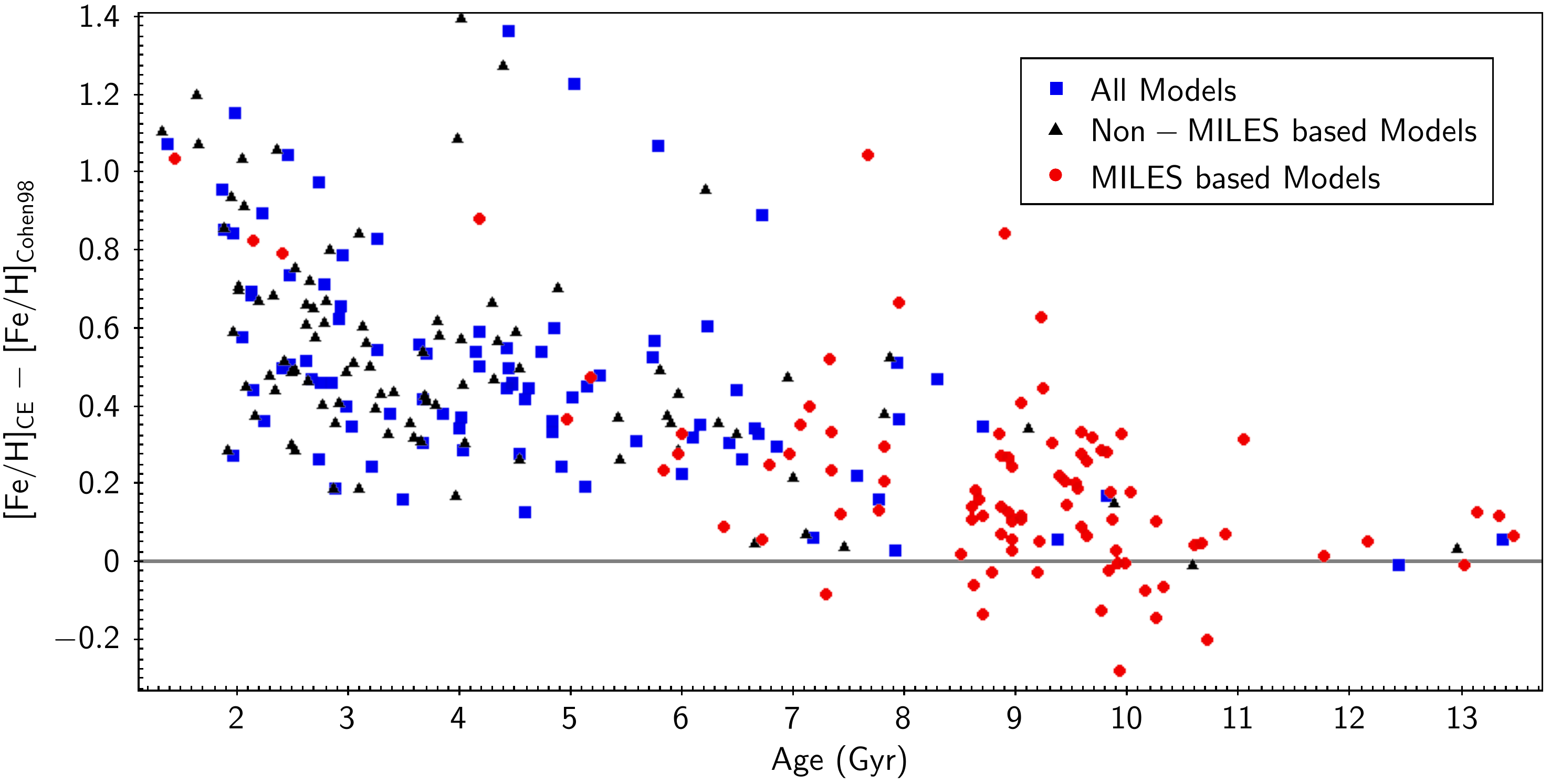}
\caption[]{Comparison between the photometric metallicities of this paper ([Fe/H]$_{{\rm CE}}$) and the spectroscopic metallicities of \citet{cohen1998} ([Fe/H]$_{{\rm Cohen98}}$) as a function of the photometric CE ages (Age$_{{\rm CE}}$). The blue squares, black triangles, and red circles are refering to the different set of models used to determine the photometric ages and metallicities (see Section\,\ref{inf_lib}). The grey line is set to 0. The trend which appears in this plot reflects the well-known age metallicity degeneracy. The MILES based-estimates seem to agree the most with the spectroscopic estimates of \citet{cohen1998}. However, it has to be kept in mind that the spectroscopic estimates have been derived with a forced age around 12 Gyr.}
\label{fig:appendix}
\end{center}
\end{figure*}

Only few databases of spectroscopic ages or metallicities for extragalactic GCs exist in the literature.
For M87 GCs, a catalog of spectroscopic metallicities was published by \citet{cohen1998}.
It has 92 objets in common with our sample.
Figure\,\ref{fig:appendix} compares the metallicities inferred by \citet{cohen1998} ([Fe/H]$_{{\rm Cohen98}}$), with those derived with the CE of our paper ([Fe/H]$_{{\rm CE}}$) for the three subsets of SSP models described in Section\,\ref{inf_lib}).
It appears that the set with all the models (blue squares) and the one with the Non-MILES models (black triangles) predict higher metallicities and lower ages than the spectroscopic estimates of \citet{cohen1998}.
The relationship between inferred age and inferred metallicity is driven by the well-known age metallicity degeneracy. Conversely, the CE based on the MILES-library models (red circles) are showing reasonable agreement with [Fe/H]$_{{\rm Cohen98}}$ ($\pm 0.2$\,dex). 
Although these two sets of estimates agree in terms of metallicity values (assuming that the [Fe/H] reference is the same), one should keep in mind that the spectroscopic metallicities presented in \citet{cohen1998} have been derived using a forced age of 12 Gyr. If we repeat our analysis with an imposed age of 12 Gyr, the three sets of models give CEs of metallicity consistent with the spectroscopic estimates. Prior assumption on age play a strong role in photometric and spectroscopic studies of remote clusters. 

\bibliographystyle{apj.bst}  
\bibliography{biblio_paper2} 

\end{document}